\documentclass[usenatbib,preprint]{mn2e}

\usepackage {natbib,aas_macros}
\usepackage {natbib,aas_macros}
\usepackage{graphicx}
\usepackage {bm}
\usepackage {amsmath,amssymb}

\title{Formation and Early Evolution of Circumstellar Disks in Turbulent Molecular Cloud Cores}

\author[Tsukamoto \& Machida]{Yusuke Tsukamoto$^{1,2}$ and Masahiro N. Machida$^{3}$ \\
$^1$Department of Astronomy, the University of Tokyo, Hongo 7-3-1, Bunkyo-ku, Tokyo, Japan\\
$^2$Division of Theoretical Astronomy, National Astronomical Observatory of Japan, 
2-21-1 Osawa, Mitaka, Tokyo, Japan\\
$^3$Department of Earth and Planetary Sciences, Kyushu University, 
6-10-1 Hakozaki, Higashi-ku, Fukuoka, Fukuoka, Japan
}

\begin{document}
\bibliographystyle{mn2e}
\maketitle

\begin{abstract}
We investigate the formation and evolution of circumstellar disks in turbulent cloud cores 
until several $10^4$ years after protostar formation
using smoothed particle hydrodynamics (SPH) calculations. 
The formation and evolution process of circumstellar disk in turbulent cloud cores differs substantially 
from that in rigidly rotating cloud cores. 
In turbulent cloud cores, a filamentary structure appears before the protostar formation and the protostar forms
in the filament. If the turbulence is initially  sufficiently strong, the remaining filament twists around the protostar 
and directly becomes a rotation-supported disk. 
Upon formation, the disk orientation is generally misaligned with the angular momentum of 
its host cloud core and it dynamically varies during the main accretion phase, even though the turbulence is weak. 
This is because the angular momentum of the entire cloud core is mainly determined by the large scale 
velocity field whose wavelength is comparable to the cloud scale, whereas the angular momentum  of the disk is determined by 
the local velocity field where the protostar forms and these two velocity fields do not correlate with each other.
In the case of disk evolution in a binary or multiple stars, the disks are misaligned with each other 
at least during the main accretion phase,
because there is no correlation between the velocity fields around the position where each protostar 
forms. In addition, each disk is also misaligned with the binary orbital plane. 
Such misalignment can explain the recent observations of misaligned disks and misaligned protostellar outflows. 
\end{abstract}


\section{Introduction}
\label{sec:intro}
Many circumstellar disks are observed in star-forming regions. 
They are by-products of star formation and are directly connected to planet formation.
Thus, it is important for understanding star and planet formation to
clarify the formation and evolution of circumstellar disks. 
Recent studies on protostar and disk formation in a collapsing molecular cloud core suggest 
that the circumstellar disk forms at a very early phase of star formation 
\citep{1998ApJ...508L..95B,2009MNRAS.400...13W,2010ApJ...724.1006M,2010ApJ...718L..58I,2011MNRAS.417.2036B}. 
and is more massive than the protostar for at least $10^4$ years 
after protostar formation \citep{2010ApJ...718L..58I}. During this phase, the circumstellar disk 
is gravitationally unstable and develops non-axisymmetric spiral arms.
However, these studies have examined the formation and evolution of the circumstellar 
disk in the rigidly rotating cloud core.

Molecular clouds, which typically have a scale $>0.1$ pc, usually 
display complex internal motions that are observed as a broad emission-line profile.
The internal motion in a molecular cloud is believed to be caused by turbulence. 
Although a detailed generation process of the internal motions is still unclear, observed 
line profiles in molecular clouds are consistent with Gaussian velocity fields with a 
Kolmogorov spectrum \citep{1995ApJ...448..226D,2000ApJ...535..869K}.
Observations also indicate the existence of turbulence even for molecular cloud cores, 
which typically  have a scale of 0.1-0.01 pc. \citet{1993ApJ...406..528G} showed that the specific angular 
momentum of a molecular cloud core is roughly proportional to  
$j \propto R^{1.6}$, which is derived from  observations of molecular cloud cores with a size of $0.06-0.6$ pc. 
This scaling law is in good agreement with the scaling, $j \propto R^{1.5}$ which is produced by a turbulence 
field with velocity power spectrum, $P(k) \propto k^{-4}$.
Note that when the molecular cloud core rotates rigidly,
 the specific angular momentum should obey the scaling law of $j \propto R^{2}$. 
\citet{2000ApJ...543..822B} pointed out that random Gaussian velocity fields with power 
spectra of $P(k)\propto k^{-4}$ can reproduce the observed projected 
rotational properties of molecular cloud cores.

The evolution of a circumstellar disk during Class 0--I phases may be 
strongly affected by the velocity field of the molecular cloud core because it is mainly 
determined by mass accretion from the (turbulent) envelope.
Thus, it is important to investigate the disk evolution in turbulent molecular cloud cores for 
a comprehensive understanding of the evolution of circumstellar disks.

In addition to formation and evolution of the circumstellar disk around a single protostar, 
cloud turbulence is also expected to affect the 
formation and evolution of circumstellar disks in binary systems. 
In a collapsing cloud core, fragmentation may occur, causing a binary system 
to appear. In a rigidly rotating cloud core,
disks in the binary system are aligned with each other by definition and also with the binary orbital plane 
\citep[see,][]{2011MNRAS.416..591T}. On the other hand, 
observational evidence indicates misalignment of binary disks. 
For example, \citet{1994ApJ...437L..55D} observed protostellar jets with different orientations in a binary system. 
\citet{2011A&A...534A..33R} recently observed two disks in a proto-binary system. 
In their observation, the primary protostar has
an almost face-on disk, whereas the secondary disk is edge-on toward the observer. 
Thus, the disk orientation around the primary protostar is almost 
perpendicular to the orientation of the disk around the secondary protostar. With observations of many 
wide binary systems, \citet{1994AJ....107..306H} showed that the stellar 
rotational equatorial  plane is often misaligned with the binary orbital plane. 
These observations indicate that in wide binary systems, the disk plane is frequently misaligned 
with the binary orbital plane. 
However, simulations starting from a molecular cloud core with simple systematic rotation cannot
explain these observations.

The gravitational collapse of a turbulent molecular cloud core has been studied by several groups.
\citet{2011ApJ...728...47M} calculated the gravitational contraction of turbulent cloud cores and 
showed the morphological evolution of the collapsing cloud core before protostar formation.
With long term calculations (until $\sim 10^5$ years after the protostar formation), 
\citet{2004A&A...423..169G,2004A&A...414..633G} showed that fragmentation frequently occurs 
even in a weakly turbulent environment. 
Recently, \citet{2010MNRAS.402.2253W} calculated the evolution of turbulent cloud cores and 
showed that protostar and circumstellar disk formation in turbulent cloud cores is different from that 
in rigidly rotating cloud cores. 
They showed that protostars form from a filamentary structure and pointed out that the 
formation condition for a binary or multiple stellar 
system rarely depends on the total angular momentum of the host cloud core. 
\citet{2012MNRAS.419..760W} investigated the evolution of low-mass cold cores. 
They focused on the relationship between the evolution of the core and
the maximum wavelength (or minimum wavenumber) of the turbulence and
found that the maximum wavelength of the turbulence is a critical parameter for the evolution of cloud cores.
In their simulation, the turbulent energy is fixed and the dependence of the disk evolution on the turbulent energy is
still unclear.

In this study, we investigate circumstellar disk formation in a turbulent cloud core over $>10^4$ years. 
We have already reported the disk formation in cloud cores with  systematic rotation 
(or rigid rotation) in \citet{2011MNRAS.416..591T}
in which we calculated the cloud evolution with different classical cloud parameters \citep[e.g.,][]{1984ApJ...279..621M}
 representing the cloud thermal ($\alpha$) and rotational ($\beta$) energies.
In the present study, which is complementary to that of \citet{2011MNRAS.416..591T},
we investigate the effects of turbulence strength on the evolution of the protostar 
and circumstellar disk with one cloud parameter representing thermal ($\alpha$) energy and one representing 
turbulent ($\gamma_{\rm turb}$) energy.
We also investigate disk evolution and its orientation in a binary system.
This paper is organized as follows. In \S 2, we describe the numerical method and initial conditions. In \S 3,
we present the numerical results.  Finally, in \S 4, we discuss our results.

\begin{table*}
\begin{center}
\caption{Model parameters}		
\begin{tabular}{cccccccc}
\hline\hline
 Model & $\alpha$ & $\gamma_{\rm turb}$ & {\it R} (AU) & $\beta_{\rm eff}$  & $\rho_{\rm init} ~({\rm g~cm^{-3}})$  & Mach number$^1$ & Fragmentation \\
\hline
 1  & 0.6 & 0.1  & 5900 & $1.3 \times 10^{-2}$ & $6.9\times 10^{-19}$ & 0.66 & N\\
 2  & 0.6 & 0.06 & 5900 & $7.9 \times 10^{-3}$ & $6.9\times 10^{-19}$ & 0.50 & N\\
 3  & 0.6 & 0.03 & 5900 & $4.1 \times 10^{-3}$ & $6.9\times 10^{-19}$ & 0.36 & N\\
 4  & 0.4 & 0.3  & 3933 & $4.1 \times 10^{-2}$ & $2.3\times 10^{-18}$ & 1.4  & N\\
 5  & 0.4 & 0.1  & 3933 & $1.4 \times 10^{-2}$ & $2.3\times 10^{-18}$ & 0.80 & N \\
 6  & 0.4 & 0.06 & 3933 & $8.5 \times 10^{-3}$ & $2.3\times 10^{-18}$ & 0.63 & N\\
 7  & 0.2 & 0.3  & 1967 & $4.0 \times 10^{-2}$ & $1.9\times 10^{-17}$ & 2.0  & Y\\
 8  & 0.2 & 0.1  & 1967 & $1.3 \times 10^{-2}$ & $1.9\times 10^{-17}$ & 1.1  & Y\\
 9 & 0.2 & 0.06 & 1967 & $8.2 \times 10^{-3}$ & $1.9\times 10^{-17}$ & 0.87 & Y\\
\hline
\end{tabular}
\end{center}
$^1$ Mach number is the root mean square of the turbulent velocity in the initial cloud core.
\end{table*}

\section{Numerical Method and Initial Conditions}
\label{method}
\subsection{Numerical Method}
Our simulations were conducted using the smoothed particle hydrodynamics (SPH) code, 
which we used in our previous study \citep{2011MNRAS.416..591T}. The code includes an 
individual time-step technique and uses the Barnes-Hut tree algorithm to calculate the self-gravity with an 
opening angle $\theta=0.5$. We include an artificial viscosity according as prescribed
by \citet{1997JCoPh.136..298M} with $\alpha_v=1$ and also use the Balsara switch \citep{1995JCoPh.121..357B}. 
Our code was parallelized with MPI and was verified using several standard test problems.

To mimic the thermal evolution of the cloud core calculated by \citet{2000ApJ...531..350M}, 
we adopted the following barotropic equation of state,
\begin{eqnarray}
\label{eos}
P=c_{s,0}^2\, \rho \left[1+\left(\frac{\rho}{\rho_c}\right)^{2/5} \right],
\end{eqnarray}
where $c_{s,0} = 190\, {\rm m~ s^{-1}}$ and $\rho_c = 4 \times 10^{-14}\, {\rm g~ cm^{-3}}$. 

In addition, to calculate the evolution of the circumstellar disk for $\gtrsim10^4 $ years, 
we adopted the sink particle technique described by \citet{1995MNRAS.277..362B}. 
We assume that a protostar forms when the particle density exceeds the 
threshold density, $\rho_{\rm sink}=4\times 10^{-9} {\rm g~ cm^{-3}}$. Next, a sink particle with an
accretion radius of 1 AU is dynamically introduced. 

\subsection{Initial Settings}
As the initial state, we adopt a spherically-symmetric cloud core with an isothermal temperature of $T=10$\, K. 
Each cloud core has a uniform density within the range $\rho_{\rm init} = 6.9\times10^{-19}$ to 
$1.9\times10^{-17}\,{\rm cm}^{-3}$ and a size ranging from $R=1967$ to $5900$\,AU (see, Table~1). 
All models have the same cloud mass of $1 M_\odot$. 
The initial cloud cores are modeled with about 520000 SPH particles. 
The mass resolution of all calculations is $1.9\times 10^{-6}~M_{\odot}$
and our calculations fulfill the resolution requirement suggested by \citet{1997MNRAS.288.1060B}.
The smoothing length of i th particle, $h_i$ is given as 
\begin{eqnarray}
h_i=1.2 \left( \frac{m_i}{\rho_i} \right)^{1/3},
\end{eqnarray}
where $m_i,~rho_i$ are the mass and the density of i th particle.
Only a turbulent velocity field without the systematic rotation velocity is imposed on the initial cloud core.

\subsubsection{Model Parameters}
Gas accretion from the infalling envelope onto 
the circumstellar disk controls the evolution of the circumstellar disk. The gas accretion rate 
is related to the thermal and kinetic (or turbulent) energies of the host cloud core. 
In this study, to investigate disk evolution in cloud cores with different thermal 
and turbulent energies, we use two parameters, $\alpha$ and $\gamma_{\rm turb}$. The parameter 
$\alpha$ is the ratio of  thermal energy ($E_{\rm thermal}$) to 
gravitational energy ($E_{\rm grav}$) in the initial cloud core and given as
\begin{equation}
\alpha = \frac{E_{\rm thermal}}{|E_{\rm grav}|}=\frac{5R_0c_{s,0}^2}{2GM},\\
\end{equation}
where $R_0$ and $M$ are the initial radius and mass of the cloud core, respectively. 
The parameter $\gamma_{\rm turb}$ is the ratio of turbulent  
energy ($E_{\rm turb}$) to  gravitational energy ($E_{\rm grav}$):
\begin{equation}
\gamma_{\rm turb} = \frac{E_{\rm turb}}{|E_{\rm grav}|}.
\label{eq:Eturb}
\end{equation}
In equation~(\ref{eq:Eturb}), $E_{\rm turb}$ and $E_{\rm grav}$ are numerically 
estimated in the initial cloud core.
The strength of the turbulent velocity field in the initial cloud core can be described by the parameter $\gamma_{\rm turb}$.

\subsubsection{Turbulence Realization}
\label{sec:realization}
\citet{2000ApJ...543..822B} showed that random Gaussian velocity fields with power spectra of $P(k)\propto k^{-3}$ to $k^{-4}$ 
can reproduce the observed projected rotational properties of molecular cloud cores. 
Thus, a molecular cloud core apparently has a systematic rotation even when it has only a turbulent velocity field. 
We adopt a turbulent velocity field with power spectra of $P(k)\propto k^{-4}$, in which we assume 
a divergence-free velocity field. The prescription for realizing turbulence in our 
initial settings is as follows: First, we generate a random Gaussian field 
with the power spectrum $P_A(k)\equiv <|A_k|^2> \propto k^{-6}$.
Next, we compute the Fourier transform of the velocity field $\bm{v_k}$ as
\begin{eqnarray}
\label{eos}
\bm{v_k}=i \bm{k} \times \bm{A_k}.
\end{eqnarray}
Finally, the velocity field is generated on a $128^3$ uniform grid, 
and the particle velocities are interpolated from the grid.
We adopted the minimum wave number to be $k_{\rm min}=1$ and the maximum wave number to be $k_{max}=128$.
Unlike a rigidly rotating cloud core, the turbulent velocity field is not uniquely identified by 
the parameter set, $(\alpha,~\gamma_{\rm turb})$ because of its stochastic nature. 
Thus, it is  difficult to systematically investigate the evolution of the turbulent 
cloud core with parameters  $\gamma_{\rm turb}$ and $\alpha$. 
\citet{2004A&A...414..633G} and \citet{2012MNRAS.419..760W} resolved 
this difficulty by considering the ensemble of simulations.
Although their approach  may be advantageous for investigating 
the turbulent cloud core, it requires too many computational resources.
Therefore, we adopted a different approach for this study in which 
we simulate the evolution of the turbulent cloud core having a "typical" 
angular momentum for each parameter set. 
To construct a velocity field with typical angular momentum for a given parameter set $(\alpha,~\gamma_{\rm turb})$, 
we generated 1000 different velocity fields for each parameter set with different random number seeds. 
Next, from the 1000 velocity fields,  we selected the velocity field whose angular momentum 
was the closest to the mean value of the angular 
momentum and used it to calculate the cloud evolution.
This procedure makes it possible  to simulate a turbulent cloud core that has a plausible angular 
momentum for a given parameter $(\alpha,~\gamma_{\rm turb})$. 
To relate the turbulent energy to the rotational energy of the initial cloud core, 
we introduce the parameter $\beta_{\rm eff}$ as
\begin{eqnarray}
\beta_{\rm eff} &=& \frac{25}{12}\frac{J_{\rm cloud}^2}{GM^3R},
\end{eqnarray}
where $J_{\rm cloud}$ is the total angular momentum of the cloud core.
This parameter can be regarded as the effective ratio of rotational energy to gravitational energy of the cloud core. 
The values of $\beta_{\rm eff}$ for each model are listed in Table 1. According to this definition, 
the effective rotational energy is roughly related to the turbulent energy as $\beta_{\rm eff} \sim 0.1\gamma_{\rm turb}$.

Using the parameters $\alpha$ and $\gamma_{\rm turb}$, 
we constructed 9 models and calculated the evolution of the
circumstellar disk for each model. 
The model names and parameters are listed in Table~1.

 Note that we adopted the turbulent velocity field but the uniform
density field as the initial state.
Thus, the velocity field and density field of our initial conditions are not self-consistent. 
We will discuss how this inconsistency would affect our results in \S \ref{inconsistency}.

Note also that some models have somewhat unrealistic parameter. 
Model 7 have large initial Mach number that is much
larger than typically observed and model 8 and 9 is highly gravitationally
unstable ($\alpha+\gamma_{\rm turb}= 0.26, 0.3$).
We will discuss the validity of these initial conditions in \S \ref{validity_alpha02}.

\begin{figure*}
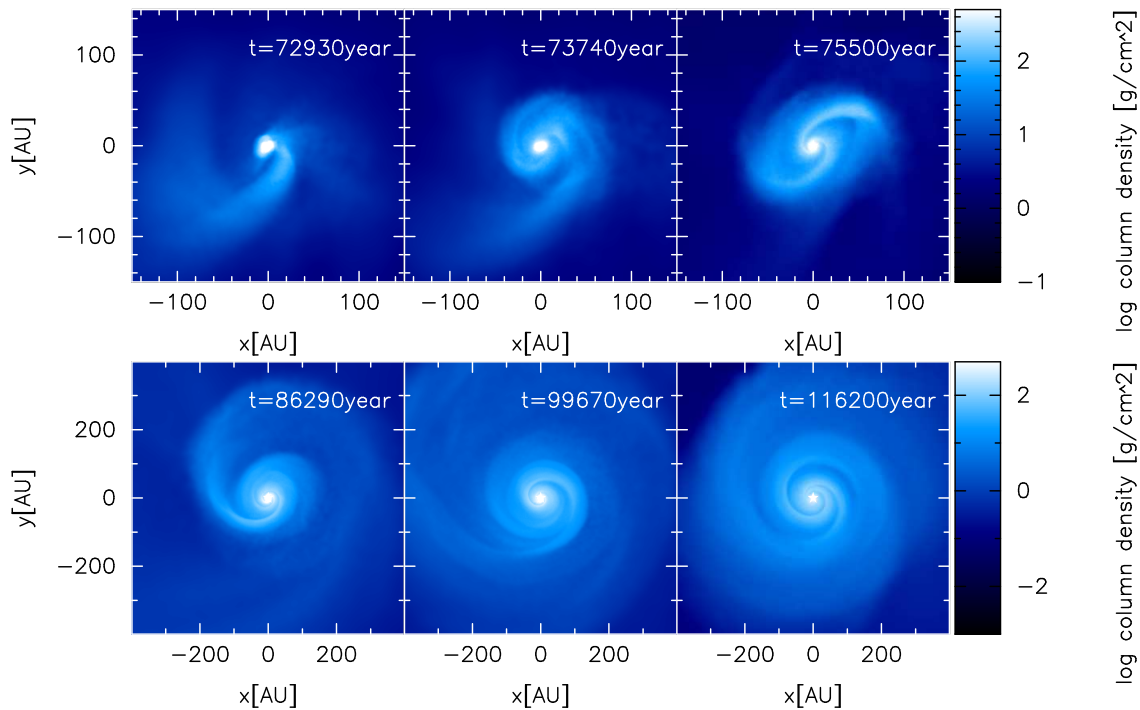

\begin{center}
\includegraphics[width=150 mm]{f1a.eps}
\includegraphics[width=150 mm]{f1b.eps}
\caption{Time sequence of the logarithm of the face-on surface density before and after protostar formation for model 4 ($\alpha=0.4$ and $\gamma_{\rm turb}=0.3$). The $z$-axis (i.e., the line-of-sight direction) is set parallel to the angular momentum of the entire initial cloud core. The top-left and -middle panels show snapshots approximately $2.6\times 10^3$ and $1.8\times 10^3$ years before protostar formation, respectively. The top-right panel shows a snapshot just when the protostar forms. The bottom left, -middle, and -right panels show snapshots $1.1\times 10^4,~2.1\times 10^4$, and $4.1\times 10^4$ years after protostar formation, respectively. The elapsed time in the calculation is shown in each panel.}
\label{faceon_strong_turbulent}
\end{center}
\end{figure*}

\begin{figure*}
\begin{center}
\includegraphics[width=80 mm,angle=-90]{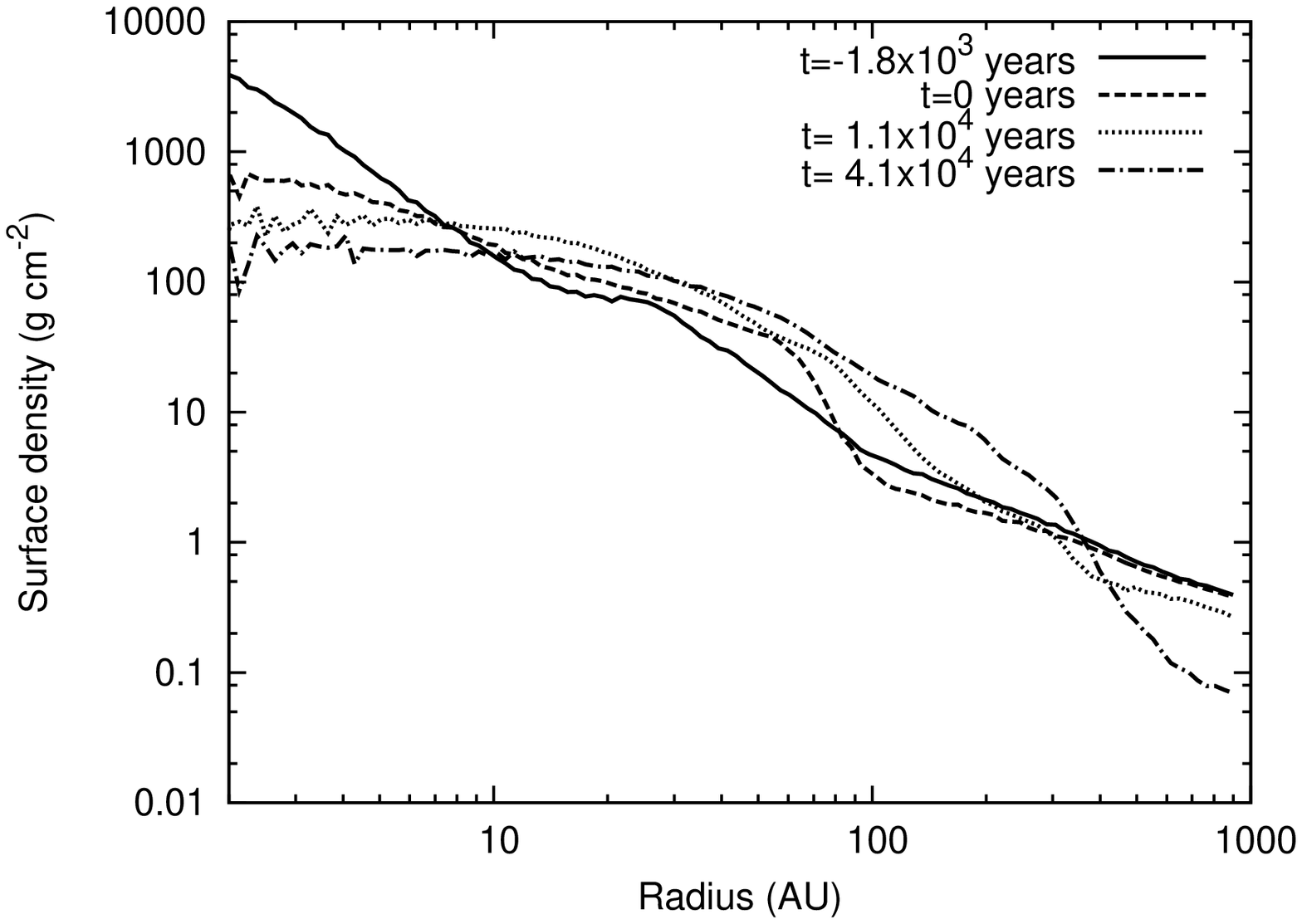}
\includegraphics[width=80 mm,angle=-90]{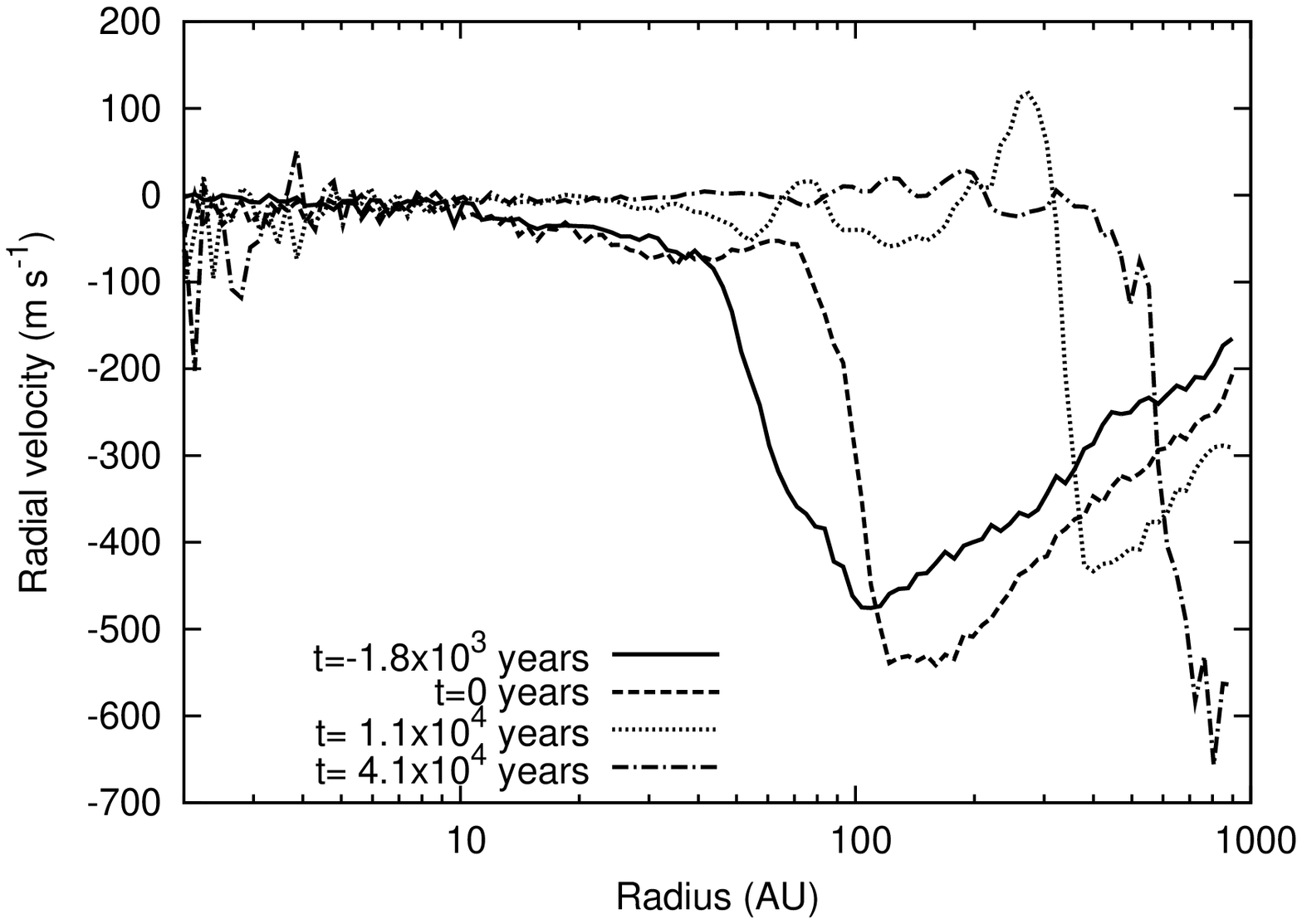}
\caption{Radial distribution of surface density (top) and radial velocity (bottom) for model 4.
Lines corresponds to $1.8\times 10^3$ years (solid line) before protostar formation, $0$ years (dashed line), $1.1\times 10^4$ years (dotted line), and $4.1\times 10^4$ years (dashed-dotted line) after protostar formation.
}
\label{radial_profile_strong_turbulent}
\end{center}
\end{figure*}

\section{Results}
\label{results}
In this section, we first give an overview of  evolution process of circumstellar disk in turbulent core and
how it changes in response to the strength of turbulence. We adopt model 4 ($\alpha=0.4$ and $\gamma_{\rm turb}=0.3$) as  
a typical model with strong turbulence and model 6 ($\alpha=0.4$ and $\gamma_{\rm turb}=0.06$) as a 
typical model with weak turbulence, and investigate them both in detail. 
It is observed that whether the initial velocity field is supersonic ($M>1$) or subsonic ($M<1$) is not important for
the evolution processes.
Next, we investigate the evolution of the disk orientation for all models and, 
finally, we investigate the formation process of binary or multiple systems in a turbulent
cloud core and the evolution process of circumstellar disks in a binary system.

\begin{figure*}
\begin{center}
\includegraphics[width=100 mm,angle=-90]{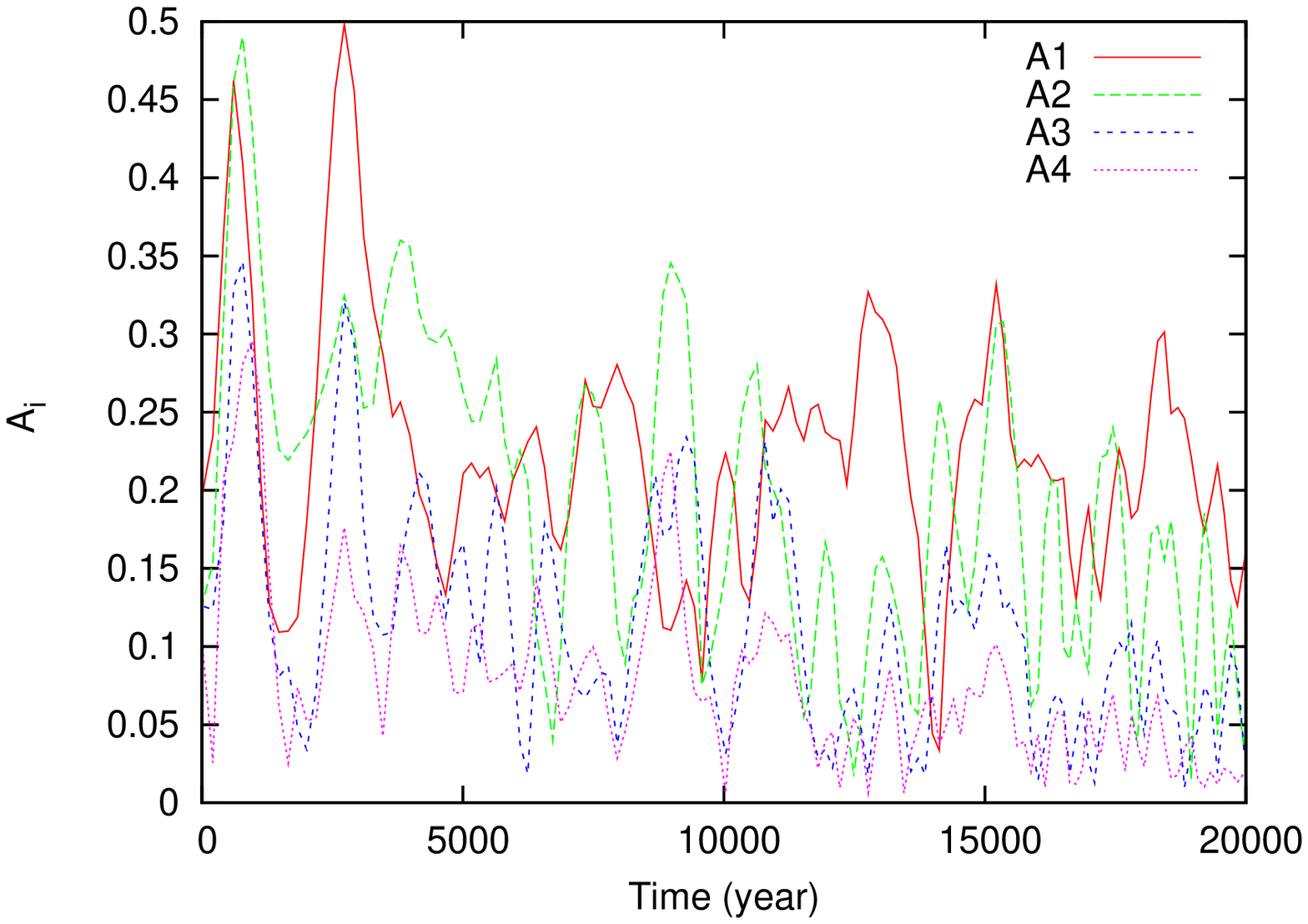}
\caption{
Evolution of Fourier amplitudes of surface density at 100 AU against elapsed time after protostar formation for model 4.
}
\label{mode_strong_turbulent}
\end{center}
\end{figure*}

\subsection{Overview}
\subsubsection{Disk evolution in strong turbulent core}
\label{sec:strong}
In this subsection, we show how the circumstellar disk evolves in a turbulent cloud core that initially 
has relatively strong turbulence. 
As a typical case, we chose model 4 with parameters $\alpha=0.4$ and $\gamma_{\rm turb}=0.3$ 
and an initial root mean square (rms) Mach number of 1.4.
Figure~\ref{faceon_strong_turbulent} shows the time evolution of the surface density around the center of the 
cloud core. In the figure, the $z$-axis is chosen to be parallel to the 
angular momentum  of the entire initial cloud core. 

Before protostar formation, the first (adiabatic) core \citep{1969MNRAS.145..271L,2000ApJ...531..350M}  appears.
The central region of high-density gas in the top-left and -middle panels of 
Figure~\ref{faceon_strong_turbulent} corresponds to the first core. These panels show that filamentary 
structure accompanies the first core, which then, coils around the first
core and changes directly to a circumstellar disk. 
 The second collapse occurs at $t=75500$ years
(top-right panel). By this epoch, the two spiral arms already
develop. 
Although we cannot recognize the filamentary structure in the top-right panel, a weak large-scale filament remains 
around the disk at this epoch.
In addition, the gas accretion from the filamentary structure onto the disk continues.
After protostar formation, this large-scale filament (the structure which resides at $\sim200$\,AU) 
coils around the disk and directly connects to the disk's spiral arm (bottom-left panel).
Finally, as seen in the bottom-right panel of Figure~\ref{faceon_strong_turbulent}, 
the filamentary structure accretes onto the disk and disappears.

Previous studies, which assumed the rigidly rotating cloud core with large rotational energy 
($\beta$ of the rigidly rotating core is comparable to $\beta_{\rm eff}$ of the turbulent core), 
have shown that the first core has enough angular momentum to  develop a
bar-mode instability and trailing spiral arms appears. The spiral arms remove
the angular momentum from the central region. With the angular momentum transfer, the central density 
becomes high and the second collapse occurs.
Then, the remnant of the first core which increases in size with the spiral arms becomes a circumstellar disk.
\citep{1998ApJ...508L..95B,2011MNRAS.417.2036B,2011MNRAS.413.2767M}.
However, the disk formation mechanism seen in Figure~\ref{faceon_strong_turbulent} differs 
from that found in the previous studies. 

The first core formed in the turbulent cloud core has less angular momentum than that 
formed in the rigidly rotating cloud core 
because the gravitational collapse tends 
to occur at a stagnation point of the velocity field 
around which the rotational energy of the gas is relatively 
low. 
As a result, the first core cannot develop the bar-mode instability even with the strong turbulence.
Instead, the filamentary structure, which forms before the first core formation, twists around the 
first core or protostar and, then, becomes directly a rotationally supported disk. 

Figure~\ref{radial_profile_strong_turbulent} shows the radial profiles of the surface density and the radial
velocity, for which each value is azimuthally averaged.
The origin is set at the position of the protostar 
(or the position of the gas particle that has maximum density before the second collapse). 
The velocity is measured in the standard of rest of the protostar. The solid line in the top 
panel shows the surface density profile $\sim1.8 \times 10^3$ years before the second collapse, 
which corresponds to the top-middle panel of Figure \ref{faceon_strong_turbulent}. At this epoch, 
the disk surrounding the first core ranges from $10$ to $50$\,AU. As described above, 
the filamentary structure, which formed before protostar formation, changes directly into the disk. 
The dashed line in the top panel of Figure~\ref{radial_profile_strong_turbulent} is the surface 
density profile at which the second collapse just begins. The disk radius reaches $\sim100$\,AU at this epoch. 
The dotted line shows the surface density profile at $\sim1.1\times10^4$ years after 
the second collapse. At this epoch, as seen in the bottom panel of Figure~\ref{radial_profile_strong_turbulent}, 
a positive radial velocity arises at approximately $200$ to $300$\,AU. 
This flow is caused by the strong $m=1$ mode of the  
spiral arm which is generated approximately 100\,AU. This type of outward flow appears 
repeatedly and readjust the surface density toward the stable configuration.

As shown in Figure~\ref{faceon_strong_turbulent}, the structure of the disk is  highly non-axisymmetric, which 
is caused by non-axisymmetric accretion from large-scale filaments. 
To clarify this, we plot in Figure~\ref{mode_strong_turbulent} the Fourier amplitudes of the 
surface density at 100\,AU for $2.0\times 10^4$ years 
after protostar formation.
The Fourier amplitudes are calculated from a Fourier series,
\begin{eqnarray}
\frac{\Sigma(R,\phi)}{\bar{\Sigma}(R)}=\sum_{\rm m=1}^{\infty}A_{\rm m}\exp[-i{\rm m}\phi],
\end{eqnarray}
where $m$ is the azimuthal wavenumber and $A_m$ is the Fourier amplitude. Figure~\ref{mode_strong_turbulent} shows 
that most of the time, the $m=1$ mode dominates the other modes, 
although the $m=2$ mode occasionally dominates the $m=1$ mode. In addition, 
the amplitude of other higher modes ($m\ge3$) is smaller than that of modes $m=1$ and 2. 
The prominent $m=1$ mode is a significant feature of disks with non-axisymmetric envelope accretion.

\begin{figure*}
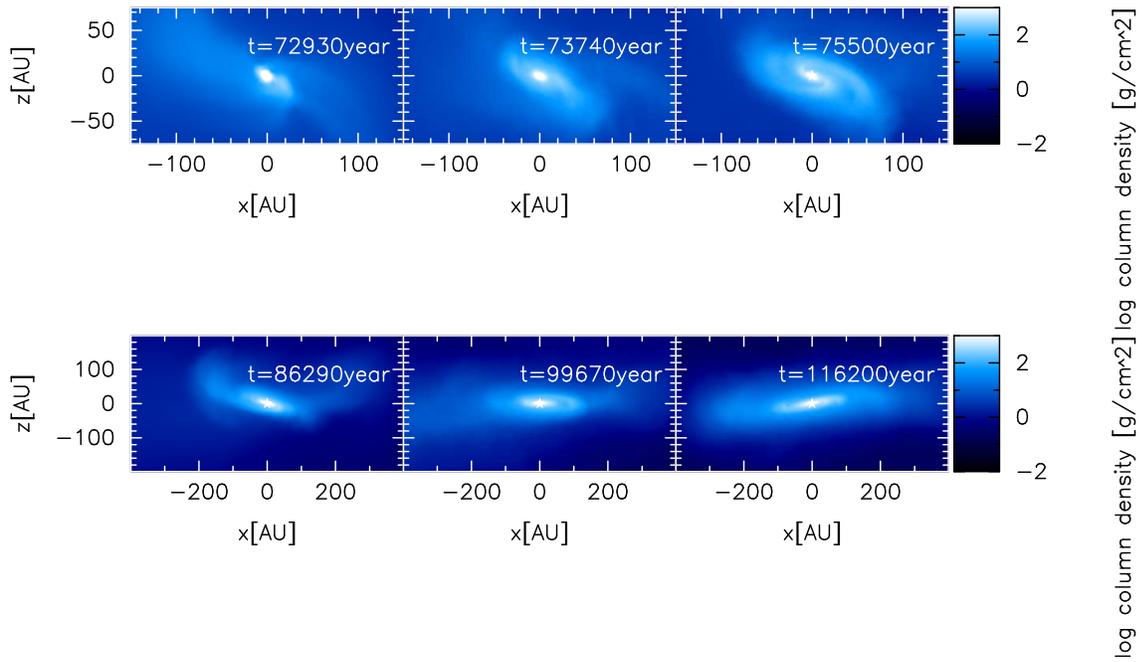

\begin{center}
\includegraphics[width=150 mm]{f2a.eps}
\includegraphics[width=150 mm]{f2b.eps}
\caption{
Same as Figure~\ref{faceon_strong_turbulent} but edge-on view. 
}
\label{edgeon_strong_turbulent}
\end{center}
\end{figure*}

Figure \ref{edgeon_strong_turbulent} shows the time evolution of an edge-on view of the center of 
the cloud core for model 4. The figure indicates that the disk orientation varies during main accretion phase. 
We discuss the evolution of the disk orientation further in \S \ref{sec:inclination}.

As we showed above, the evolution process of a circumstellar disk in a strong turbulent core 
is different from that in a rapidly rotating core in many respects. As we will
see below, the evolution process becomes similar to that of  a rigidly rotating cloud core as the 
turbulent energy decreases. However, note that the dynamical change of the disk orientation also occurs even in
the weak turbulent core.

\begin{figure*}
\begin{center}
\includegraphics[width=150 mm]{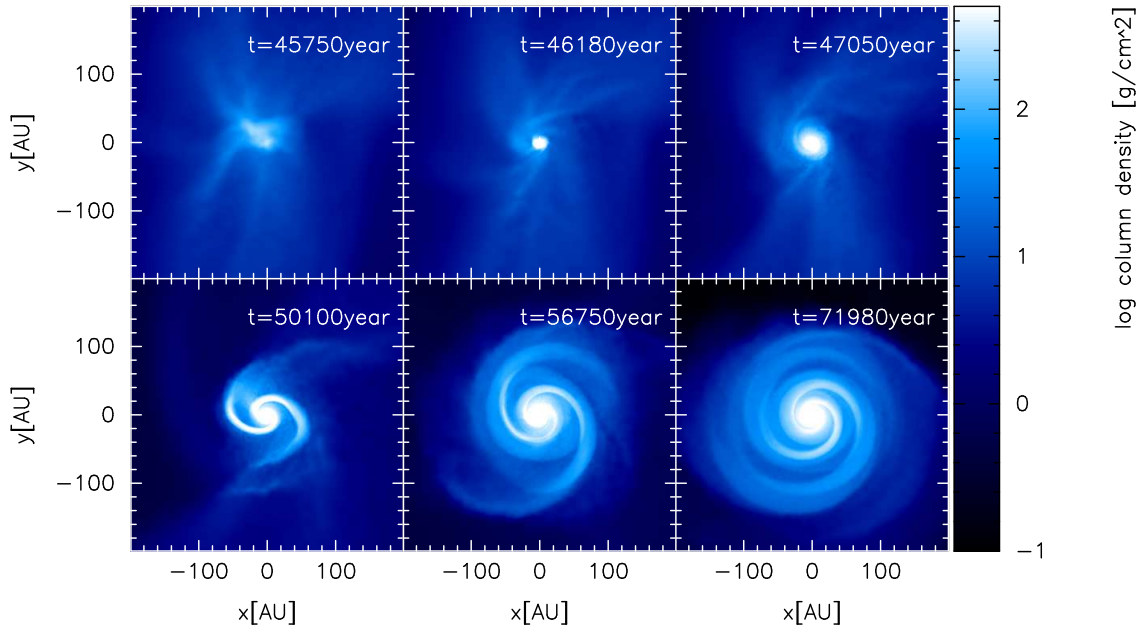}
\caption{
Time sequence of logarithm of face-on surface density before and after protostar formation for model 6 
($\alpha=0.4$ and $\gamma_{\rm turb}=0.06$). The $z$-axis is parallel to the angular momentum of the entire initial cloud core. 
Top-left panel shows a snapshot approximately $4.3\times 10^2$ years before protostar formation. Top-middle panel shows a snapshot just when the protostar forms. Top-right, bottom-left, -middle, and -right show snapshots $8.7\times 10^2$\,years, $3.9\times 10^3$\,years, 
$1.1\times 10^4$\,years, and $2.6 \times 10^4$ years after protostar formation, respectively. The elapsed time in the calculation is shown in each panel.
}
\label{faceon_weak_turbulent}
\end{center}
\end{figure*}

\subsubsection{Disk evolution in weak turbulent core}
As a typical case for the formation and evolution of a circumstellar disk in a weakly turbulent cloud core, we chose model 6 
that has parameters of $\alpha=0.4$ and $\gamma_{\rm turb}=0.06$. The initial rms Mach number is 0.63. 
Compared with model 4 (strong turbulent case), model 6 has the same thermal energy but a smaller turbulent energy.

\begin{figure*}
\begin{center}
\includegraphics[width=150 mm]{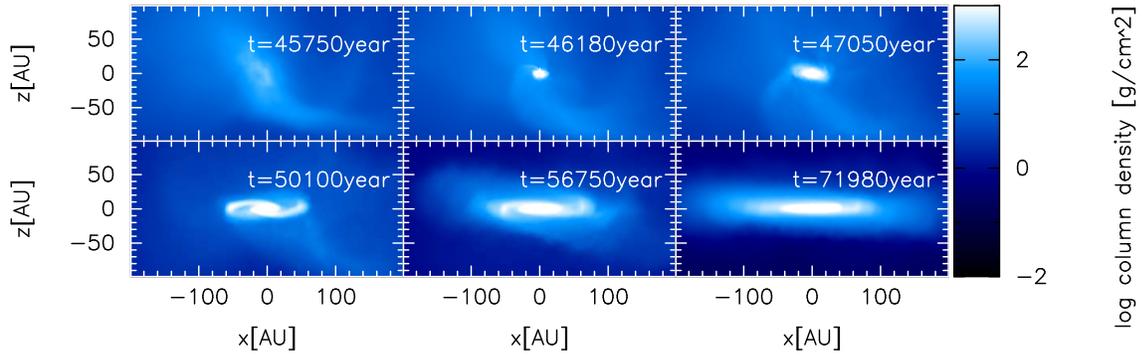}
\caption{
Same as Figure~\ref{faceon_weak_turbulent} but edge-on view. }
\label{edgeon_weak_turbulent}
\end{center}
\end{figure*}

\begin{figure*}
\begin{center}
\includegraphics[width=80 mm,angle=-90]{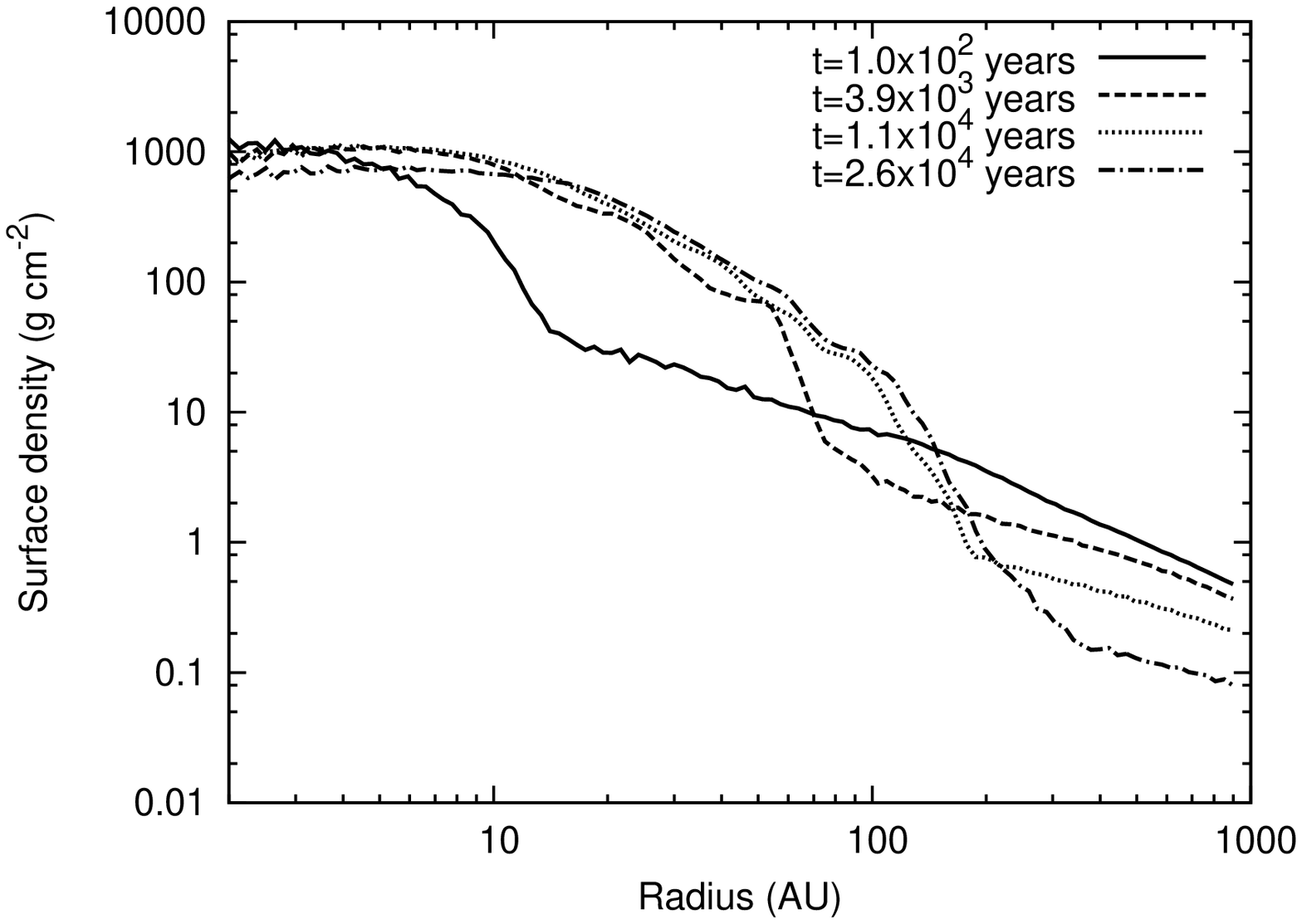}
\includegraphics[width=80 mm,angle=-90]{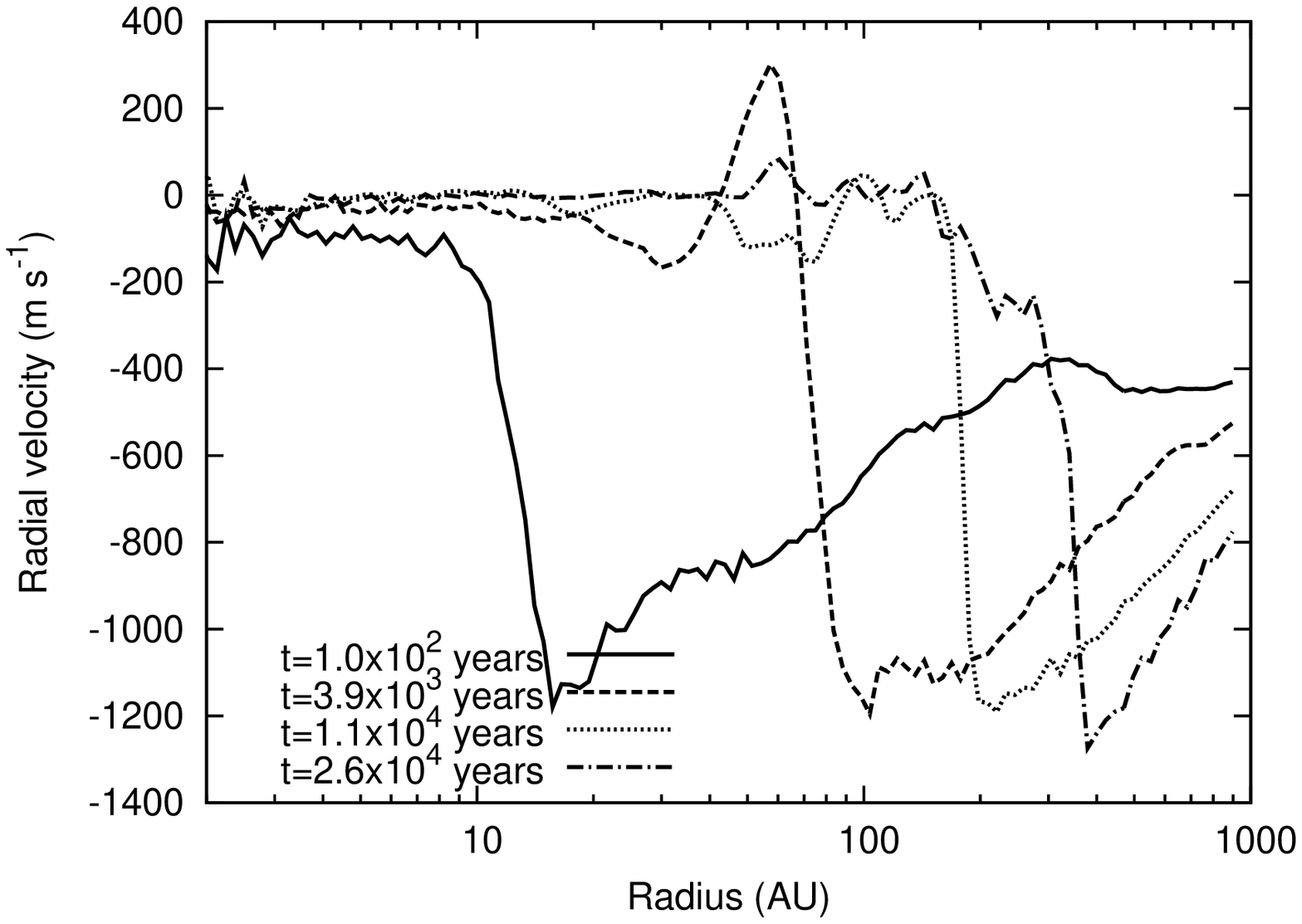}
\caption{
Radial distribution of surface density (top) and radial velocity (bottom) for model 6.
Lines corresponds to $ 1.0\times 10^2$\,years (solid), $3.9 \times 10^3$\,years (dashed), $1.1\times 10^4$\,years (dotted), and $2.6\times 10^4$ (dashed-dotted)\,years after protostar formation.
}
\label{radial_profile_weak_turbulent}
\end{center}
\end{figure*}

\begin{figure*}
\includegraphics[width=100 mm,angle=-90]{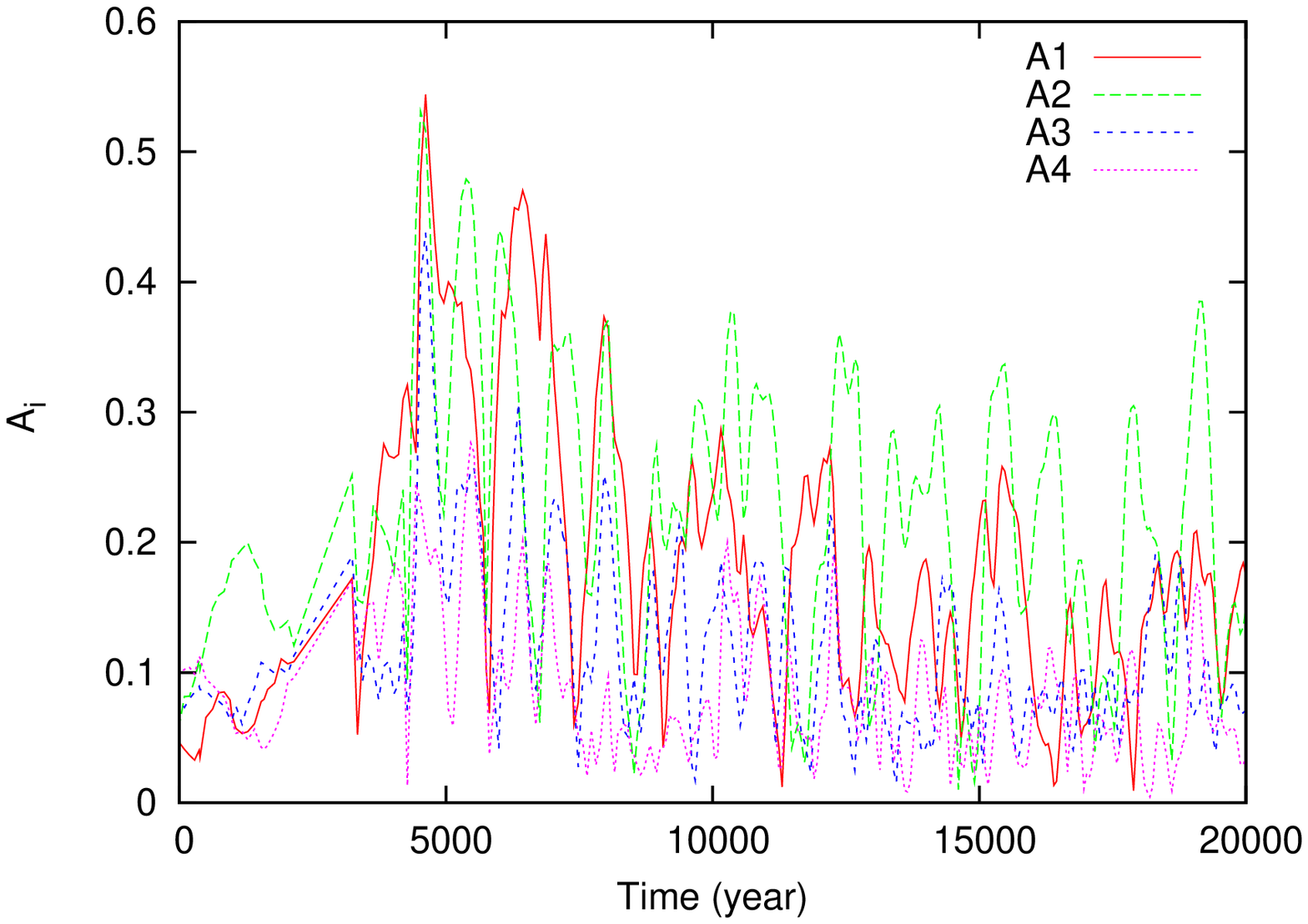}
\caption{
Evolution of Fourier amplitudes of surface density at 100 AU versus elapsed time after protostar formation for model 6.
}
\label{mode_weak_turbulent}
\end{figure*}

Figure \ref{faceon_weak_turbulent} shows a face-on view of the time evolution of the center of the 
cloud core. Before protostar formation (top-left panel), 
some filaments appear and form a complicated structure. However, the density amplitude of filaments appearing 
in this model is much weaker than what appears in the models which initially have strong turbulence. 
As the mass accretion onto disk proceeds, the circumstellar disk becomes massive and 
develops the spiral arms (bottom panels).
Finally, the radius of the circumstellar disk reaches $\gtrsim100$\,AU.
The filaments that appear in the early evolution phase (top panels) are attributed to 
anisotropic gas accretion in a weakly turbulent environment, whereas the spirals 
that appear in the later evolution phase (bottom panels) 
develop because of gravitational instability of the disk itself.

Figure~\ref{edgeon_weak_turbulent} shows the disk evolution when viewed edge-on.
The results indicate that the disk orientation varies with time even with the weak turbulence.
As we will see in \S \ref{sec:inclination}, the dynamical change of the disk orientation generally occurs independent of 
 the strength of the turbulence.
The fluctuation of the disk orientation is a significant feature of disk evolution in turbulent cloud cores.

Figure~\ref{radial_profile_weak_turbulent} shows the radial profiles of the surface density (top) 
and radial velocity (bottom) for model 6; the solid, dashed, dotted, and dashed-dotted lines 
correspond to the top-middle, bottom-left, bottom-middle, and bottom-right panels in Figure~\ref{faceon_weak_turbulent}, 
respectively. The top panel of Figure~\ref{radial_profile_weak_turbulent} shows that the disk 
gradually increases in size and surface density with time. 
In addition, the figure shows that a disk-like structure of size $\sim10$\,AU already exists before and 
immediately after protostar formation (solid line). This indicates that the remnant of the 
first core changes directly into the circumstellar disk. 
Thus, the formation and evolution processes of the 
circumstellar disk in weakly turbulent cloud cores are qualitatively the
same as those in slowly and rigidly
rotating cloud cores seen in previous works \citep[see, e.g.,][]{2011MNRAS.417.2036B,2010ApJ...724.1006M}. 
In the bottom panel of Figure~\ref{radial_profile_weak_turbulent}, a positive radial velocity 
arises in the range of 50--100 AU at $\sim5.0 \times 10^3 $ years after the second collapse. 
This outward flow in the disk is due to the $m=2$ mode of spiral arms.

Figure~\ref{mode_weak_turbulent} shows the Fourier amplitudes of the surface density at 100 AU 
for $2\times10^4$ years after protostar formation for model 6. 
For $t\lesssim 5\times10^3$ years, the disk radius 
is smaller than 100\,AU.  Thus, these amplitudes during this epoch originate 
not from the circumstellar disk but from anisotropic accretion from the infalling envelope.
After the circumstellar disk grows sufficiently, unlike the model which have strong turbulence, 
the $m=2$ mode dominates the other modes, especially for $t\gtrsim 1\times 10^3$ years, as shown in 
the bottom-right panel of Figure~\ref{edgeon_weak_turbulent}. 

As seen in previous studies \citep{1998ApJ...504..945L,2005MNRAS.358.1489L}, without gas 
accretion onto the disk, the gravitational instability of the disk tends to develop the $m=2$ spiral arms.
Because of the weak turbulence in this model, the gas falls almost isotropically 
onto the circumstellar disk. Thus, the prominence of the $m=2$ mode is 
attributed to the gravitational instability of the disk. On the other hand
, the model with the strong turbulence, the $m=1$ density perturbation mode is dominant in the 
disk. This $m=1$ mode is considered to be caused by anisotropic gas accretion 
onto the circumstellar disk. Note that the initial density perturbation of the disk strongly affects the non-linear 
development of the spiral arms \citep{1994ApJ...436..335L}.
These major modes play an important role in angular momentum transfer and 
readjust the surface density toward a more stable configuration.

\subsection{Evolution of Disk Orientation}
\label{sec:inclination}
\begin{figure*}
\begin{center}
\includegraphics[width=50 mm,angle=-90]{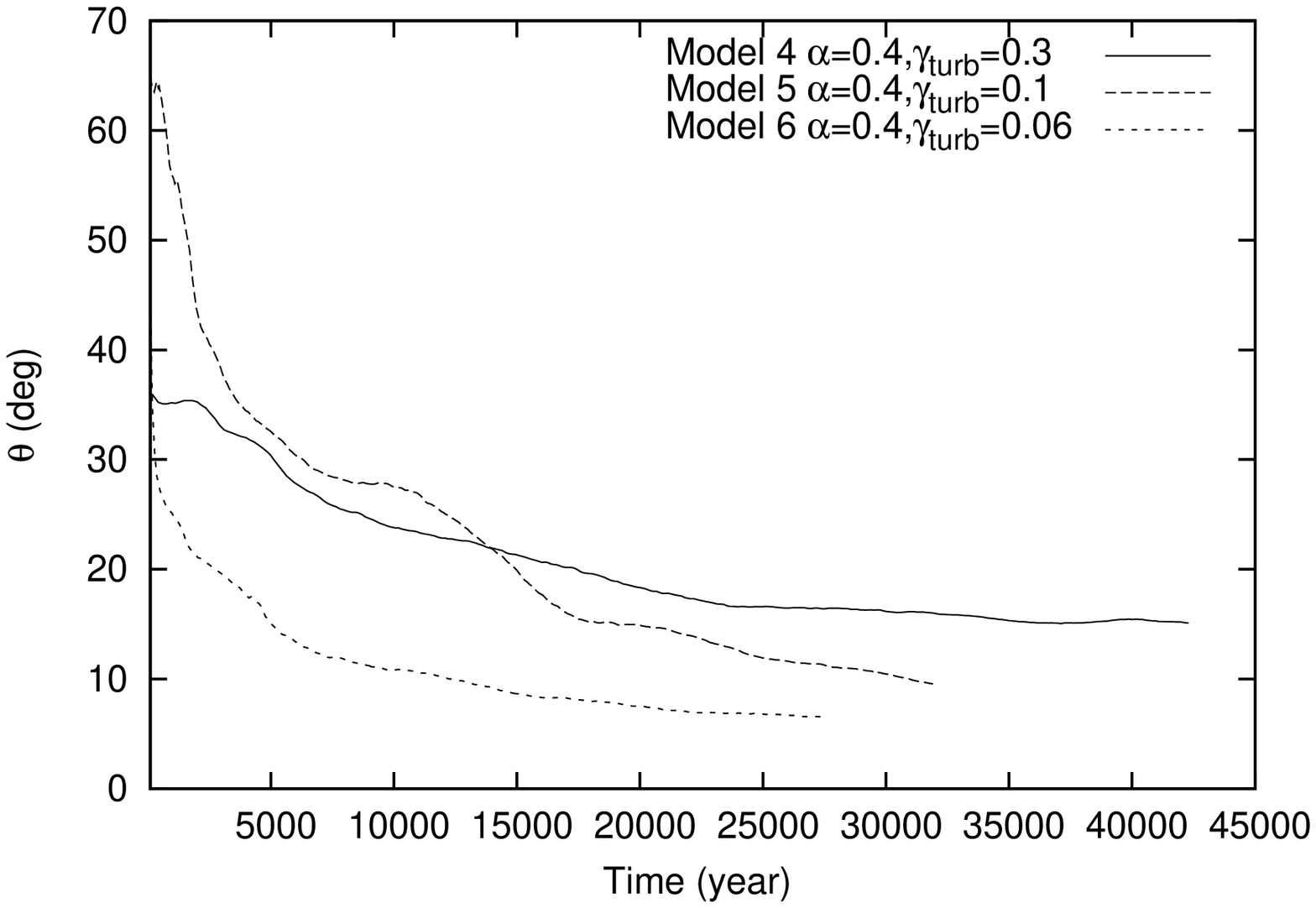}
\includegraphics[width=50 mm,angle=-90]{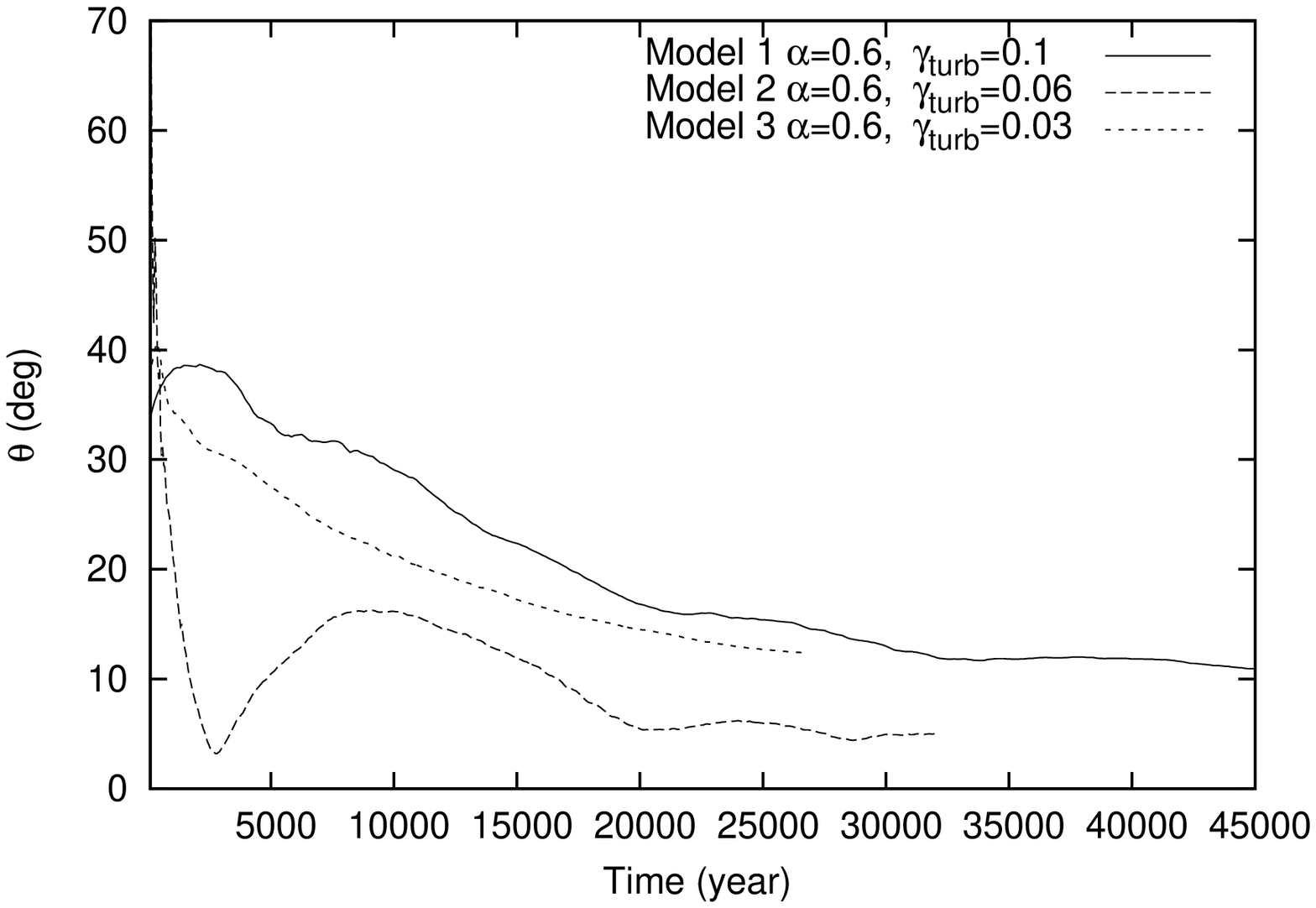}
\caption{
Time evolution of orientation of disk angular momentum. 
The orientation angle $\theta=0$ corresponds to the orientation of the angular momentum  of the initial cloud core. 
Left panel shows models having $\alpha=0.4$, in which the solid, dashed, and dotted lines correspond to models with parameters of $\gamma_{\rm turb}$ =0.3, 0.1, and 0.06, respectively. 
Right panel shows models having $\alpha=0.6$, in which the solid, dashed, and dotted lines correspond to models with $\gamma_{\rm turb}$ = 0.1, 0.06, and 0.03, respectively.
}
\label{inclination_disks}
\end{center}
\end{figure*}

Figure~\ref{inclination_disks} shows the time evolution of the angle of orientation of the disk 
angular momentum with respect to the angular momentum of the entire initial cloud core 
for non-fragmentation models (models 1--6). Because the disk is mainly supported by rotation, 
the orientation of the disk angular momentum roughly corresponds to the disk orientation. 
The disk angular momentum (disk orientation) is parallel to the angular 
momentum of the initial cloud core when $\theta=0$. In the figure, 
the origin of time is set to the epoch when the protostar forms.
This figure shows that the disk orientation is generally misaligned with respect to the total angular 
momentum of its host core at its formation and dynamically 
changes during the main accretion phase regardless of the strength of the turbulence.
In each model, the angle of orientation of the disk is $30^\circ$--$60^\circ$ toward the 
angular momentum of the initial cloud core at the protostar-formation epoch ($t=0$). 
As the accretion proceeds, the angle of orientation 
gradually decreases, reaching $\theta \lesssim 15^\circ$ within several $ 10^4$ years for all models.
It is expected that the disk orientation will be aligned with the angular momentum of the cloud core
when the gas accretion ceases.

The right panel of Figure~\ref{inclination_disks} shows that the angle of orientation
for model 2 ($\alpha=0.6$ and $\gamma_{\rm turb}=0.06$) varies irregularly with time. 
In the model, a very compact disk ($\lesssim 10$\,AU) forms at first. It accretes onto the 
protostar and disappears in a short duration of $\sim10^3$ years. 
Then, another larger disk forms by the subsequent mass accretion. Reflecting this disk disappearance, model 2 
shows an irregular evolution of disk orientation.  
This irregular evolution in the early disk evolution stage 
may be unrealistic because of artificial treatment of the accretion onto the sink particle. 
However, this treatment for sink particle does not qualitatively change our claim that the disk 
orientation angle dynamically changes with time.

\begin{figure*}
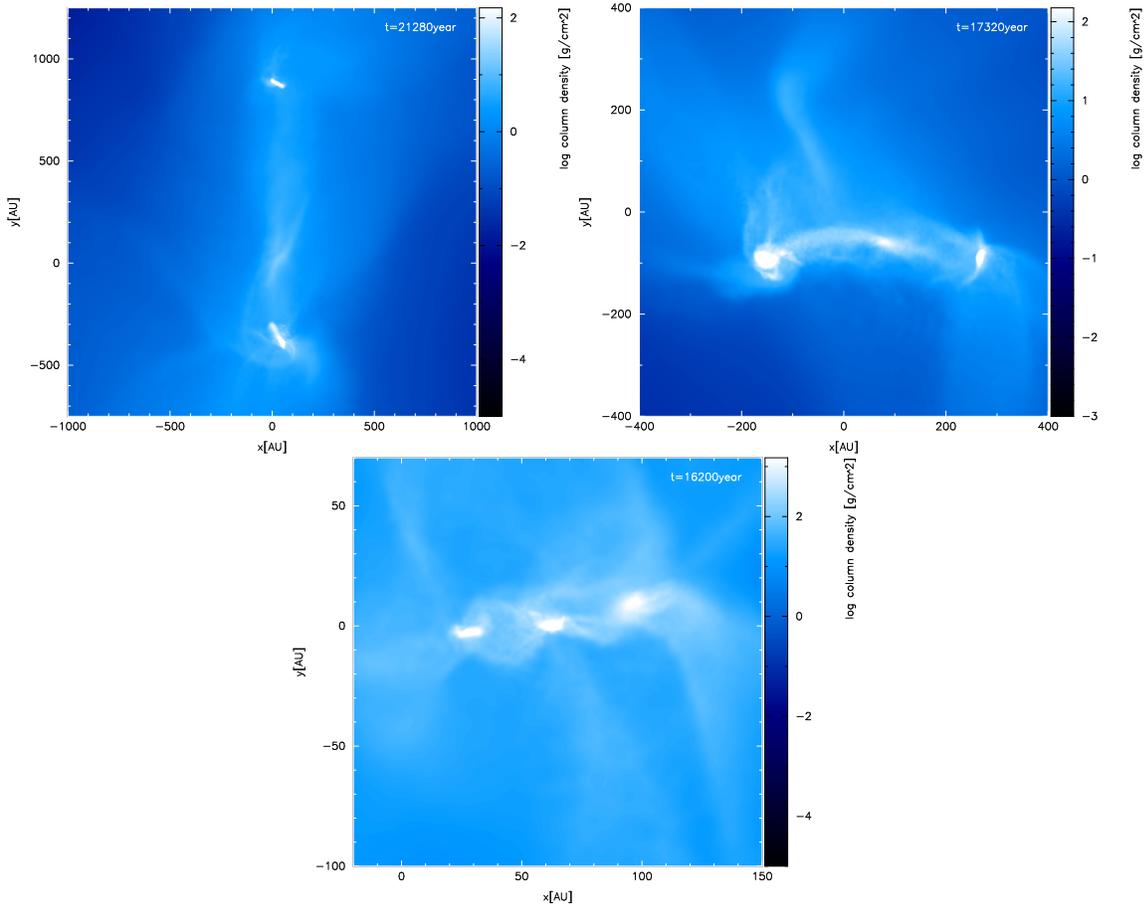

\begin{center}

\includegraphics[width=75 mm]{f7b.eps}
\includegraphics[width=75 mm]{f7a.eps}
\includegraphics[width=75 mm]{f7d.eps}
\caption{
Logarithm of face-on surface density when a binary or multiple system
 appears in  models  7 (top left, $\gamma_{\rm turb}=0.3$),  8 (top right, $\gamma_{\rm turb}=0.1$) and 9 (bottom, $\gamma_{\rm turb}=0.06$).
Models 7-- 9 have the same initial thermal energy of $\alpha=0.2$. 
}
\label{fragmentation_morphology}
\end{center}
\end{figure*}

\begin{figure*}
\begin{center}
\includegraphics[width=50 mm,angle=-90]{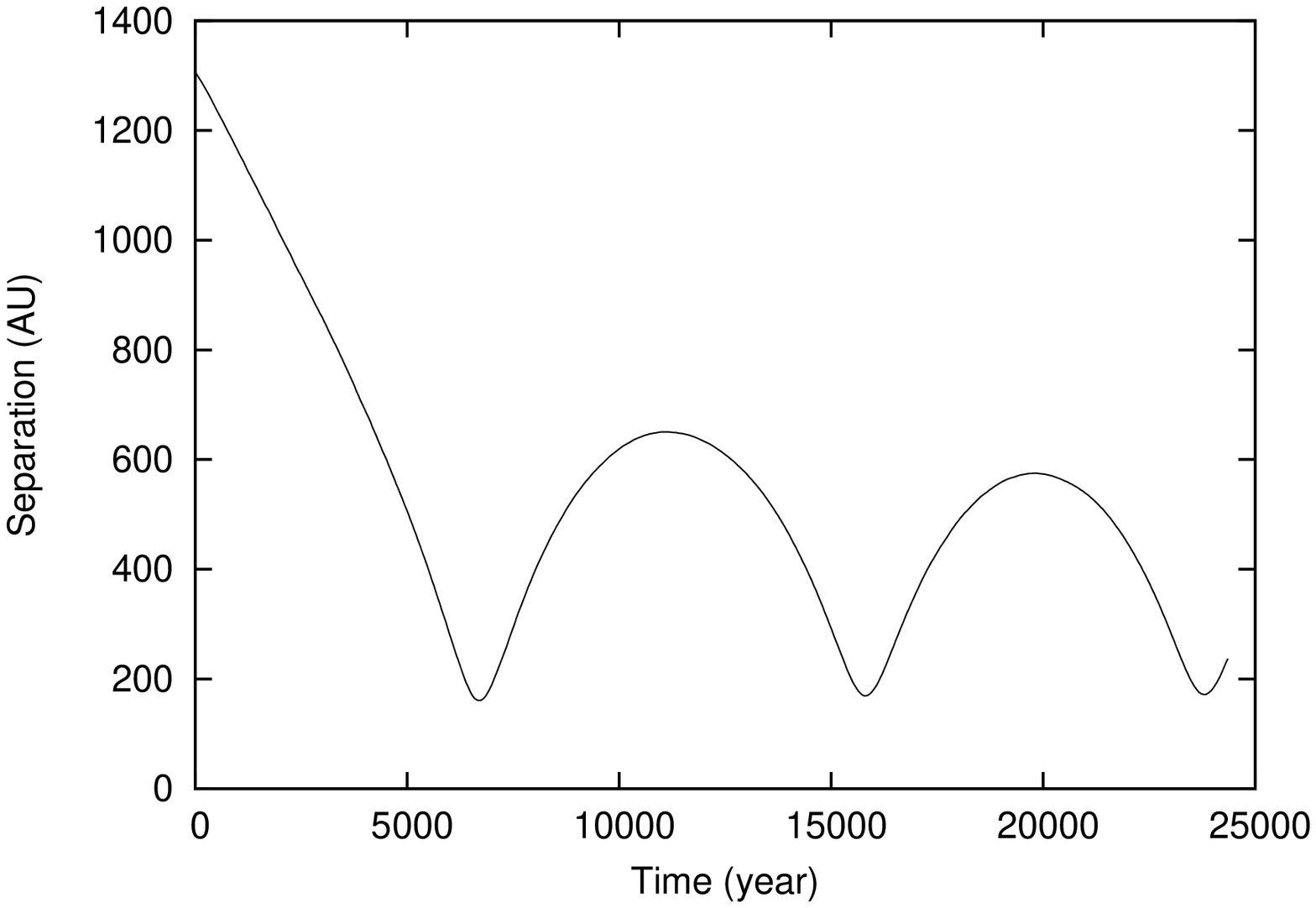}
\includegraphics[width=50 mm,angle=-90]{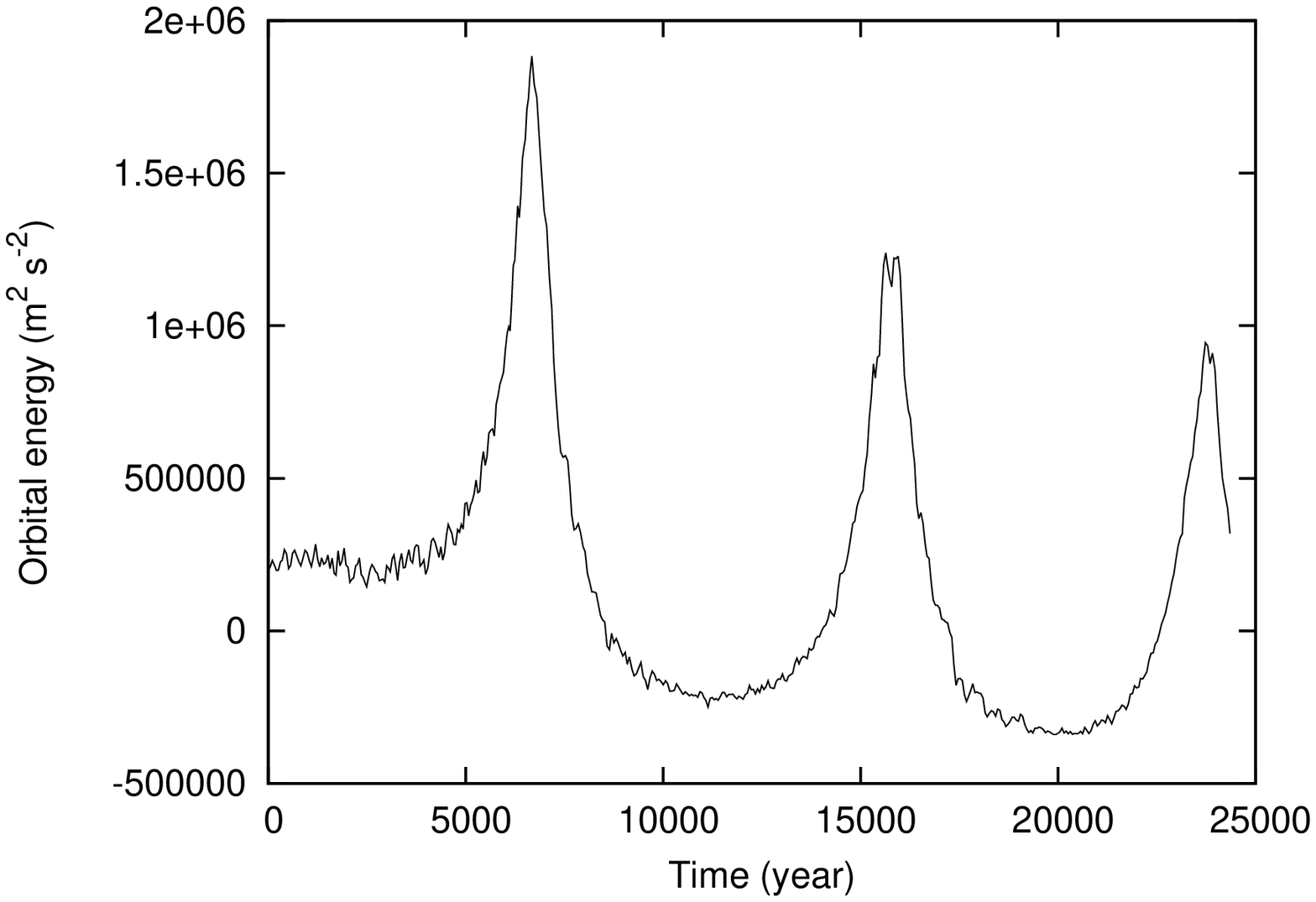}
\caption{
Time evolution of the separation between primary and secondary
 protostars (left) and the orbital energy (right) 
 for model 7.
}
\label{separation_binary}
\end{center}
\end{figure*}

\begin{figure*}
\begin{center}
\includegraphics[width=100 mm]{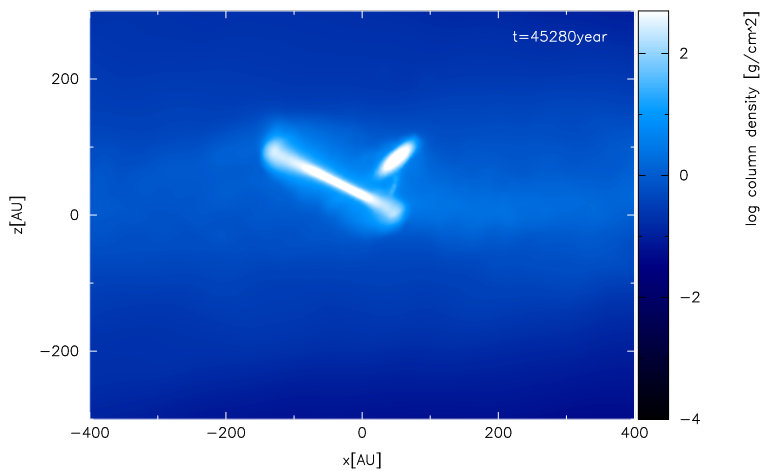}
\caption{
Surface density distribution at end of calculation for model 7 ($\alpha=0.2$ and $\gamma_{\rm turb}=0.3$). Calculation stops $2.5 \times 10^4$\,years after the secondary protostar forms.
}
\label{final_snapshot}
\end{center}
\end{figure*}

\subsection{Fragmentation and Evolution of Disks in Binary System}
\label{sec:fragmentation}
In models 7--9, fragmentation of filament  occurs and a binary or multiple stellar system appears,
whereas only a single protostar forms in models 1--6.
The fragmentation process in a turbulent cloud core differs considerably from that in a cloud 
core with initial rigid rotation. 
Figure~\ref{fragmentation_morphology} shows the density distribution
when the protostars form for models 7--9. 
The figure indicates that the binary or multiple system forms by fragmentation of the filament.
The figure also shows that the disks in a binary or multiple systems are
not aligned with each other. 
For these fragmentation models, fragmentation tends to occur near the edge of the filament. 
This fragmentation process is possible to be explained by the focal point scenario suggested 
by \citet{2004ApJ...616..288B}.

\begin{figure*}
\begin{center}
\includegraphics[width=50 mm,angle=-90]{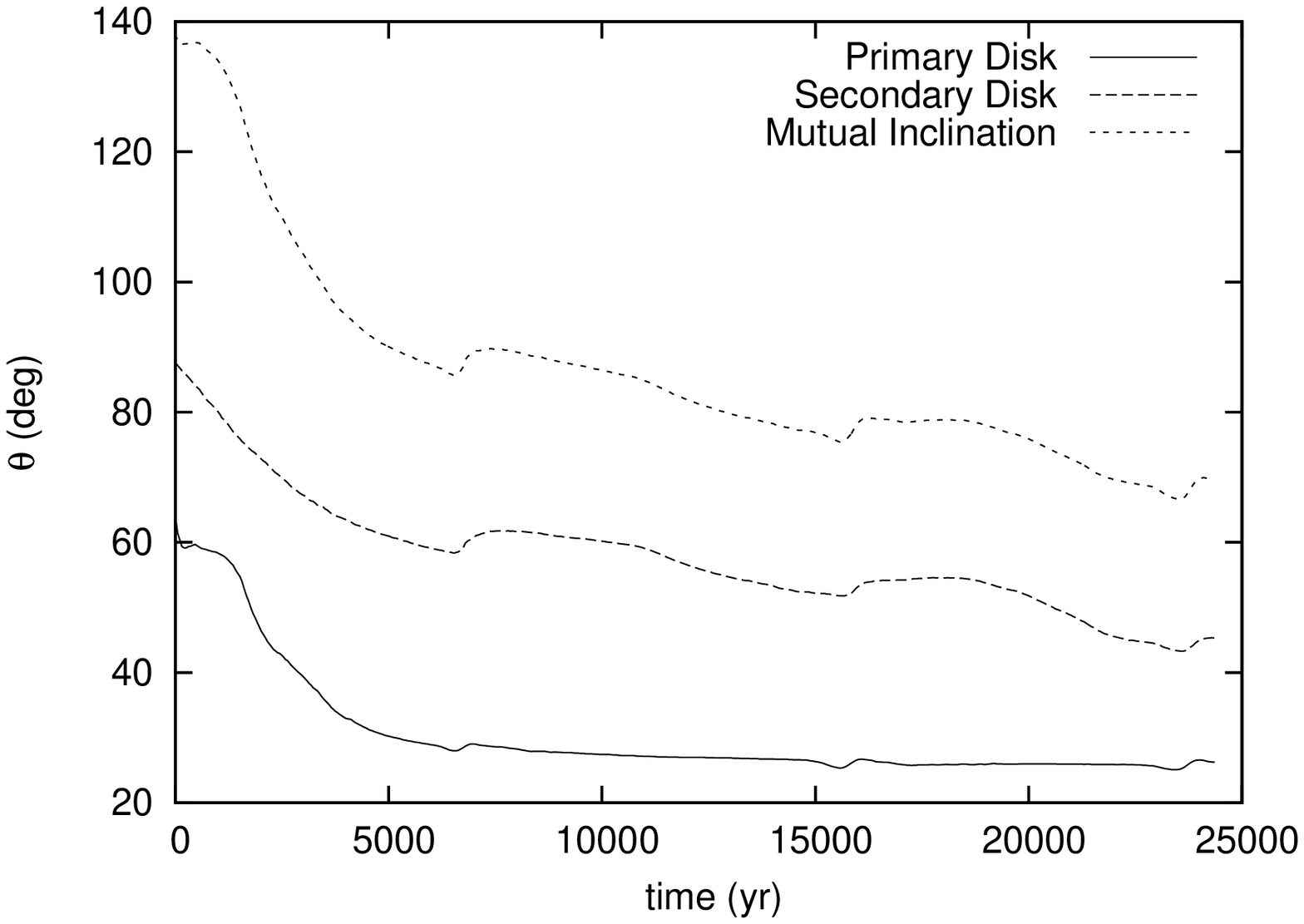}
\includegraphics[width=50 mm,angle=-90]{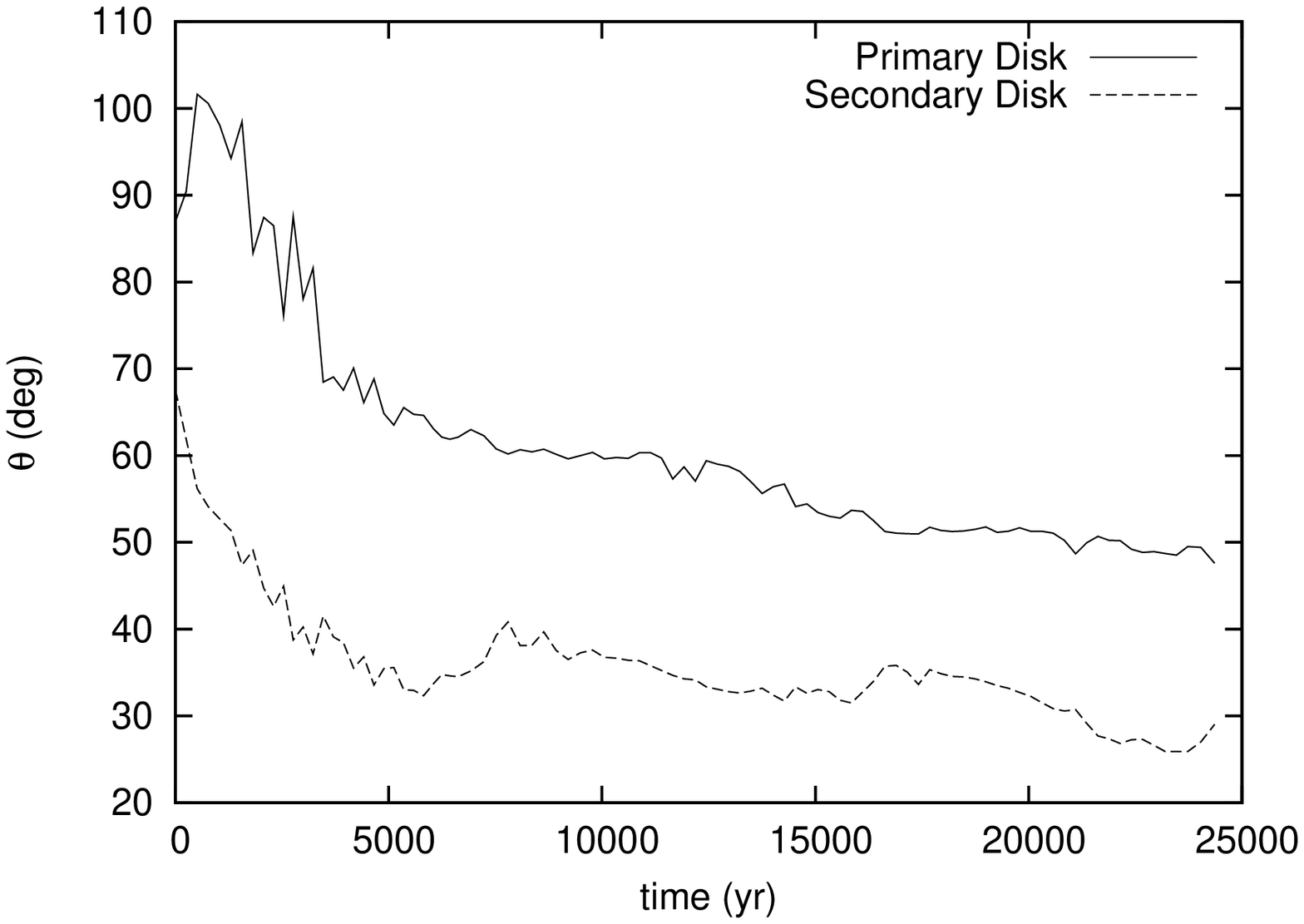}
\caption{
Left panel shows time evolution of angle of disk orientation with respect to  angular momentum of initial cloud core for primary (solid line) and secondary (dashed line) disks, and mutual inclination between the primary and secondary disks (dotted line).  
Right panel shows the time evolution of the inclination of the primary (solid line) and secondary (dashed line) disks toward the binary orbital plane.
}
\label{mutual_inclination}
\end{center}
\end{figure*}

To investigate further evolution of a wide binary system and disk evolutions, 
we focus on model 7 ($\alpha=0.2$ and $\gamma_{\rm turb}=0.3$). 
Figure~\ref{separation_binary} shows the time evolution of the binary
separation and its orbital energy for model 7. 
The orbital energy is given as
\begin{equation}
\label{orbital_energy}
  E_{\rm orbit}= \frac{1}{2} v^2 - \frac{G(m_1 + m_2)}{r}, 
\end{equation}
where $r$ and  $v$ are the relative distance and relative velocity,
respectively.
Note that the mass of protostar changes during 
the simulation. 
{\bf 
The right panel shows that the orbital energy steadily decreases,
indicating that the binary system loses its orbital energy 
by tidal interaction between the disks and stars.  In the left panel, we
can see the separation of the binary also decreases.}
Therefore, it is expected that 
the observed separation of a binary or multiple system does 
not reflect the initial separation.
{\bf In the right panel, sharp increases of the orbital energy are caused by the close encounter. 
During the early phase of binary formation, each disk is more massive than its host star
and the stars are also accelerated by their surrounding disks. 
In equation~(\ref{orbital_energy}), since the kinetic and potential
energies of the disks are not included, this orbital energy is not conserved.
Therefore, the orbital energy shows a large fluctuation at every close encounter. 
Note that it is considerably difficult to estimate the orbital 
energy of whole system which consider the stars and disks, because the disks exchange 
their mass and a part of the disk gas is striped off when close encounter occurs.
However, we can roughly understand the evolution of the
separation and the orbital energy of the binary system for a 
long duration with Figure~\ref{separation_binary}.
Note also that, the binary stars sometime have a positive orbital
energy, which seems to indicate that the binary system is not gravitationally bounded.
However, the sign of orbital energy has no meaning since we ignored the disk component in equation~(\ref{orbital_energy}).
We confirmed that the whole system (disks and stars) are gravitationally bounded during this epoch.
}

Figure \ref{final_snapshot} shows an edge-on snapshot around the binary system at the end of the simulation (
$t=2.5 \times 10^4$ years after the second protostar formation).
Because the disk orientation around each protostar is mainly determined by the local velocity field at 
which the protostar forms, the disks are not aligned with each
other. 
The misalignment is also seen in model 8.
Thus, it seems that the misalignment between disks is general feature of
binary/multiple systems which form via filament fragmentation and initially have large separation ($\gtrsim100$\,AU).

The left panel of Figure~\ref{mutual_inclination} shows the evolution of
the orientation angle of the disks from the angular momentum direction of the host core and the mutual inclination 
between the primary and secondary disks.
In this study, we call the circumstellar disk 
around the primary protostar the primary disk and that around the secondary protostar the secondary disk. 
The angle of the primary disk orientation is $\sim90^\circ$ in the very early phase of disk formation and it 
decreases with time (right panel). The sudden increases at $t=6\times10^3$, $1.6\times10^4$, 
and $2.4\times10^4$ years are due to close encounters, as seen in Figure~\ref{separation_binary}. 
At each close encounter, the angle variation for the secondary disk is larger than that for the primary disk, 
which is because of the secondary disk is less massive than the primary disk. Thus, the secondary disk is strongly affected 
by the tidal interaction.
As seen in the right panel of Figure~\ref{mutual_inclination}, the primary and secondary disks are inclined from the orbital plane by approximately $50^\circ$ and $30^\circ$ at the end of the simulations.

\section{Discussion}
\label{discussion}



\subsection{Dynamical Change of Disk Orientation}
As shown in Figure~\ref{inclination_disks}, the disk orientation is generally misaligned with the 
angular momentum of its host cloud core and dynamically changes during the main 
accretion phase even in a weakly turbulent cloud. 
This is because the angular momentum of the entire cloud core is determined mainly 
by the velocity field, whose wavelength is comparable to the cloud scale, $\lambda \sim R_{\rm cloud}$. 
On the other hand, the angular momentum of the disk is determined by the local velocity field, whose wavelength is 
much smaller than the cloud size, $\lambda \ll R_{\rm cloud}$. 
There is no phase correlation between these scales.
This nature is independent of the strength of turbulence.
The disk orientation changes due to the mass accretion which brings the angular momentum.
The variation in the disk orientation during the main accretion phase can 
explain the precessing outflow \citep[e.g.,][]{2009ApJ...692..943C}
because the outflow is believed to be driven by the circumstellar disk.

\subsection{Misaligned Disks in Binary Systems}
When large-scale fragmentation occurs and the binary system initially has a wide separation, 
disks tend to be misaligned because there is no correlation between the velocity fields 
in different regions. Thus, when a wide binary system forms in a turbulent cloud core,  
disks are expected to be misaligned with each other and also misaligned with the binary 
orbital plane at their formation epoch.
Our simulation showed that the misalignment can maintain at least for several $10^4$ years.
\citet{2011A&A...534A..33R} recently observed a proto-binary system embedded in a 
common envelope and showed that the disks are highly misaligned (edge-on and face-on disks).
They suggested that such a binary system is formed by fragmentation of two different parts of 
the collapsing molecular cloud core.  
Our study can naturally explain such misaligned disks in a binary system.  
Misaligned disks are also observed even in a multiple stellar 
system \citep{2009A&A...502..623R}. Our result also indicates that such a system can 
form via fragmentation of the collapsing cloud core in a turbulent environment, 
as seen in Figure~\ref{fragmentation_morphology}.

We can also expect the disk orientation in a binary system from observations 
of molecular outflows or optical jets. \citet{2008ApJ...686L.107C} observed 
a low-mass protostellar binary system and discovered two high-velocity 
bipolar molecular outflows that were nearly perpendicular to each other, 
showing a quadruple morphology. They concluded that the disks in a wide binary system are 
not necessarily co-aligned after fragmentation. A quadrupolar morphology and misaligned outflows are often 
observed in young binary systems \citep{1990ApJ...356..184M,2001A&A...375.1018G,2002ApJ...576..294L,2008ApJ...676.1073C}. 
Because outflows or jets are expected to be driven by the disk, misaligned disks are expected to show  misaligned outflows. 
Therefore, observations of wide binary systems and our 
results support the idea that a binary/multiple system forms via dynamical 
fragmentation of the filament in a turbulent molecular cloud core.

\subsection{Further Disk Evolution and Implications for Planet Formation in Binary System}
Tidal interaction between disks and protostars in binary system is essential for further evolution of 
disk orientation in binary system. 
\citet{1995MNRAS.274..987P} showed that perturbation of the secondary star 
can align the primary disk toward the binary orbital plane. \citet{2000MNRAS.317..773B} 
also showed that the disk can be aligned with the binary orbital plane on the precession 
time-scale owing to tidal interaction. 
Thus, whether the disk is aligned with the orbital plane depends on the parameters of 
the binary system (especially on the orbital period). 
On the other hand, observations of main-sequence binary systems indicate that the stellar rotational equatorial 
planes (which may reflect the orientations of the disks) in wide binary systems ($\gtrsim40 $\,AU) are 
often misaligned with the binary orbital plane \citep{1994AJ....107..306H}. 
Thus, in wide binary system, disk misalignment is often maintained even after the main accretion phase.

When disk misalignment remains after the main accretion phase, it is important for planet formation and 
the orbital evolution of planets in a binary system. When planet formation in a binary 
system is investigated, it is conventionally assumed that each disk 
is aligned with the binary orbital plane \citep{2000ApJ...543..328M,2007ApJ...669.1316T}.
However, when the disk and the binary orbital plane are mutually 
misaligned, the planet formation processes drastically change. 
By the Kozai effect \citep{1962AJ.....67..591K}, 
inward migration of planetesimals occurs. Thus, a compact dense planetesimal disk 
forms \citep{2011ApJ...735...10X}. On the other hand, the Kozai migration mechanism \citep{2007ApJ...669.1298F} 
seems to be preferable for explaining 
the existence of close-in planets if disk misalignment generally occurs. 
\citet{2010PASJ...62..779N} recently found that the orbital plane of the 
planets is generally not aligned with the orbital plane of the binary system. 
This also seems to be explained by the idea that the disks were 
misaligned with the binary orbital plane during the formation epoch.

\subsection{Comparison with Disk Evolution Simulations in Turbulent Cloud Core}

\citet{2004A&A...414..633G,2004A&A...423..169G} investigated the evolution of turbulent
cloud cores.
As we mentioned in \S\ref{sec:realization}, they considered an ensemble of simulations with different random number seeds, 
and  showed that fragmentation frequently occurs even with  a moderate
accretion rate that is realized in cloud cores with
initially weak turbulence ($\alpha=0.45$ and  $\gamma_{\rm turb}=0.05$). 
Their results are qualitatively the same as ours: fragmentation occurs in turbulent 
cloud cores and turbulence promotes fragmentation.
However, the fragmentation condition shown in \citet{2004A&A...414..633G,2004A&A...423..169G} seems to be quantitatively different from ours, because fragmentation only occurs with high accretion rate  ($\alpha\le0.2$ and $\gamma_{\rm turb}>0.06 $) 
in our simulations. 
This difference may originate in the difference of initial conditions.
In their study, the initial cloud cores with a Plummer-like density profile has a total mass of $M=5.4 M_{\odot}$. 
On the other hand, in our study, the initial cloud cores with a uniform density profile has a total mass of  $M=1 M_{\odot}$. 
Thus, although it is difficult to quantitatively compare our results with \citet{2004A&A...414..633G,2004A&A...423..169G}, 
the difference in initial conditions may somewhat changes the fragmentation condition. 
It is also possible that the different numerical resolutions may also affect the fragmentation condition.
They resolved the cloud core with 25000 particles, whereas we resolved it with 520000 particles. 
Mass resolutions in their study and in our study are 
$m_{\rm res}=2.1\times 10^{-4}$ and $1.9\times 10^{-6}~M_{\odot}$, respectively, which 
implies that spatial resolution in our simulation is approximately five times higher than in theirs.
As discussed in \citet{2006MNRAS.373.1039N}, there is a possibility that
the lower spatial resolution may enhance disk fragmentation. 

\citet{2010MNRAS.402.2253W} investigated the evolution of turbulent cloud cores whose mass is $M_{\rm core}=6.1 M_{\odot}$.
They fixed the mean Mach number of cloud cores to be $\mathcal{M}=1$ and changed the minimum and maximum turbulent 
wavelength and random seeds.
In their study, the sum of the thermal and kinetic energies exceeds the 
gravitational energy, (i.e., $\alpha+\gamma_{\rm turb}>1$).
Thus, initial cloud cores are gravitationally unbound as a whole. 
The cloud  begins to collapse after the energy dissipates by radiative cooling. 
By this approach, the consistency of the density distribution and the velocity field can be
realized before the collapse begins.
But it is difficult to control the accretion rate onto the disk with simple parameters.
Note that, in our study, the accretion rate can be controlled by the parameters $\alpha$ and $\gamma_{\rm turb}$ 
because we adopted cloud cores that were initially gravitationally unstable.
They pointed out that initial differences of the angular momentum rarely affects the evolution 
of the cloud and circumstellar disk.
Rather, a filamentary structure appearing in turbulent cloud cores affects the evolution of the protostar and disk. 
Their results seem to be in agreement with ours.
However, whereas a clear spiral arm often develops after disk formation in our simulation, 
most of the disks formed in their simulations does not develop the spiral arms.
This difference may come from difference in the accretion rate.

The evolution of low-mass very cold cores with a cloud mass of $M_{\rm core}=1.28~M_{\odot}$ was 
investigated in \citet{2012MNRAS.419..760W}. 
They focused on the relationship between the evolution of the core and the maximum wavelength of the turbulence.
They pointed out that the dynamical fragmentation of filaments is a major process for 
binary formation in turbulent cloud cores. 
They demonstrated that binary formation via disk fragmentation is rare in the early phase of protostar formation which 
seems to be in agreement with our results.
Their initial conditions are very cold and  turbulence is very weak. Thus,  disks formed
in their simulations are very compact compared with ours \citep[see, Fig.~\ref{faceon_strong_turbulent} 
or \ref{faceon_weak_turbulent} and Fig.~2 of ][]{2012MNRAS.419..760W}. 
In this study, we focused on the relationship between the evolution of the circumstellar disk and the turbulent energy of 
the cloud core.  Furthermore, we showed  how the disk orientation varies in time  which is not mentioned in their study.

\subsection{Validity of Initial Conditions}

\subsubsection{Inconsistency between the velocity and density
   field and possible effects on the results}
\label{inconsistency}

As described in \S \ref{method}, we imposed the turbulent velocity field
(Gaussian random field) on the uniform density field. Thus, they are
not self-consistent when the simulation begins.
Thus, it is expected that the velocity field do work to establish self-consistent density field 
during the early phase of cloud evolution and the turbulence would
decay. This decay could change the effective turbulent energy for the
simulations. 
Another possible effect is that the density field could be still
inconsistent even after the disk formation because
our initial conditions are gravitationally bounded ($\alpha +
\gamma_{\rm turb} < 1$) and gravitational collapse immediately begins.
In this subsection, we discuss whether these effects change our results.

\begin{figure*}
\begin{center}
\includegraphics[width=100 mm,angle=-90]{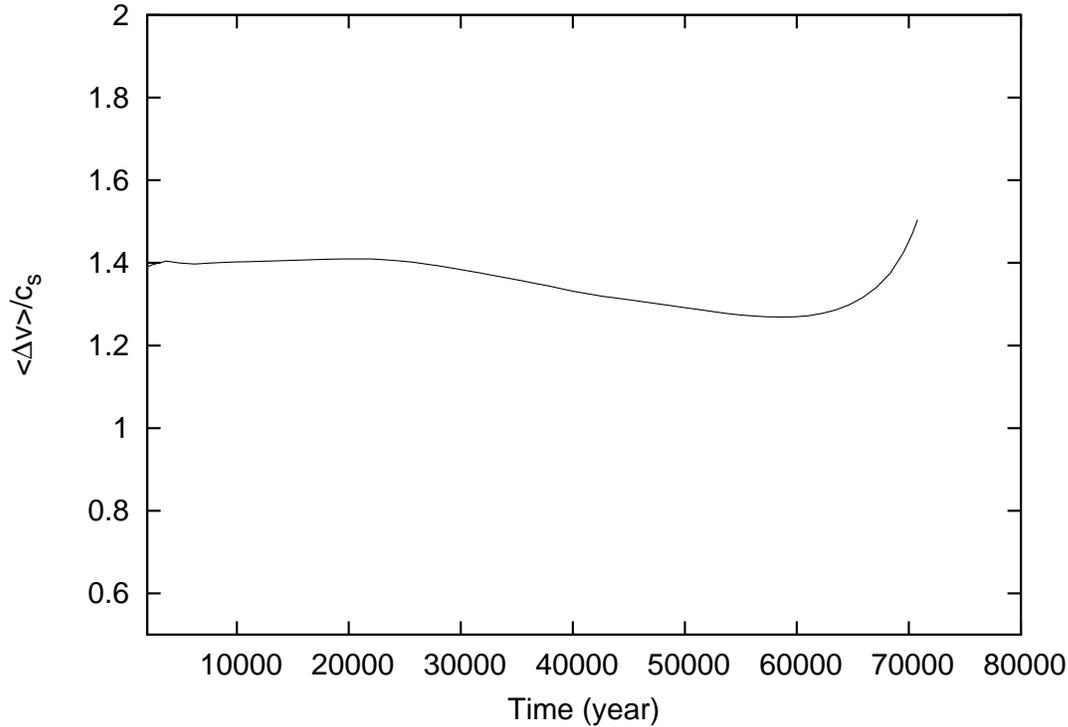}
\caption{
Velocity dispersion, $\Delta v$ as a function of time in the isothermal
 collapse phase is shown.
}
\label{mach_num_evolution}
\end{center}
\end{figure*}

In figure \ref{mach_num_evolution}, to investigate how much the turbulence decay, we show the time evolution of the 
velocity dispersion in the isothermal phase (maximum density of cloud is
less than the critical density, $\rho_{max}<\rho_c$ ) for model 4 as an example.
In the figure, the velocity dispersion slightly decreases due to the loss of turbulence energy for $t \lesssim 60000$ years. 
Then, it turns to increase at $t \sim 60000$ years, and continues to increase for $tgtrsim 60000$ years.
This increase is attributed to the infall motion of the cloud. 
This figure indicates that the decrease of turbulent energy before the cloud collapse is not so large.
Thus, it is expected that this energy loss rarely affects the subsequent cloud evolution significantly. 
We confirmed that the turbulent energy loss in other models is also small.

Next, we discuss how the inconsistent density field affects the
evolution of the system. 
In this study, we showed (i) anisotropic accretion enhances the 
low-order mode of spiral arms in the disk, 
(ii) the disk orientation dynamically changes during the main accretion 
phase, and (iii) the orientation of the disks is mutually misaligned and 
also misaligned with the orbital plane in binary system if the binary stars form in different
region  (or initial separation is sufficiently large).
They are all originated from the anisotropic gas accretion onto the disk.
The inconsistent density field would be smoother than a more realistic (or 
self-consistent) density field. Thus,
if the self-consistent density field is adopted for the initial
conditions, it is expected that the anisotropic accretion may be
enhanced
Thus, all the above features should be more stressed but not suppressed.
Therefore, our results are not changed by the inconsistent density field.

\subsubsection{Validity of High Mach Number or Highly Gravitationally
   Unstable Initial Conditions}
\label{validity_alpha02}

In this study, to comprehensively understand the relation between
turbulence and disk evolution, 
we adopted over a wide range of parameters and investigated both the
disk formation around single star and in binary system.
To investigate the disk evolution process around the binary system, we
adopted a bit artificial initial conditions:  initial cloud core of model 7 has large Mach 
number ($\mathcal{M}>1$) which is much larger than the typical observed
value in cloud cores with masses of order 1 Msolar, and cores of models 8 and 9 are in highly 
gravitationally unstable state ($\alpha+\gamma_{\rm turb}= 0.3$ and  $0.26$).
If our results and conclusions are closely related to the artificial
aspects of these initial conditions, it is problematic. 

However, we would like to emphasize that
our finding is not related to it. 
The main results we obtained from the simulations of model 7, 8 and 9
are that the disk orientation of binary system generally 
misaligned each other and from orbital plane if the initial separation
of protostars is sufficiently large. 
This is because there is no correlation between the local velocity fields where the protostars forms.
It is fundamental property of turbulence that there is no correlation
between the local velocity fields around the different positions.
Thus, such a misalignment may occur in the actual molecular cloud core in which
binary system forms. This should be confirmed by the
observation and we think the observations we mentioned above
\citep{1994AJ....107..306H,2011A&A...534A..33R} seem to support our
results.

\section *{Acknowledgments}
We thank S. Inutsuka, T. Tsuribe, K. Osuga, K. Tomisaka, S. Okuzumi, and
K. Tomida for the fruitful discussions. 
We also thank the anonymous referee for valuable comments.
The snapshots were produced by
SPLASH \citep{2007MNRAS.374.1347P}.
The computations were performed on a parallel computer, the XT4 system at
CfCA of NAOJ and SR16000 at YITP in Kyoto University. 
Y.T. is financially supported by Research Fellowships for Young Scientists from JSPS.

\bibliography{article}

\end{document}